\PassOptionsToPackage{table,dvipsnames}{xcolor}
\documentclass[hidelinks,onefignum,onetabnum,dvipsnames]{siamart250211}

\ifpdf
\hypersetup{
  pdftitle={Learning to Trust AI and Data-driven models for Data Assimilation through a Multifidelity Ensemble Gaussian Mixture Filter Framework},
  pdfauthor={A. A. Popov}
}
\fi

\usepackage[table,dvipsnames]{xcolor}
\usepackage{pgfplots}
\pgfplotsset{compat=1.18} 
\usepackage{pgfplotstable}
\usepackage{cbcolors}
\usepackage{tabularx}
\usepackage{multirow}
\usepackage{booktabs}
\usepackage{mathtools}

\usepackage{lipsum}
\usepackage{amsfonts}
\usepackage{graphicx}
\usepackage{epstopdf}
\usepackage{algorithmic}
\ifpdf
  \DeclareGraphicsExtensions{.eps,.pdf,.png,.jpg}
\else
  \DeclareGraphicsExtensions{.eps}
\fi

\newsiamremark{remark}{Remark}
\newsiamremark{hypothesis}{Hypothesis}
\crefname{hypothesis}{Hypothesis}{Hypotheses}
\newsiamthm{claim}{Claim}
\newsiamremark{fact}{Fact}
\crefname{fact}{Fact}{Facts}

\headers{Trusting AI \& data-driven models through MFEnGMF}{A. A. Popov}

\title{Learning to Trust AI and Data-driven models in Data Assimilation through a Multifidelity Ensemble Gaussian Mixture Filter Framework\thanks{Submitted to the editors \today.
\funding{This work was funded by startup fund provided by the University of Hawai'i at Manoa.}}}

\author{Andrey A. Popov\thanks{Department of Information \& Computer Sciences, University of Hawai'i at Manoa, Honolulu, HI 
  (\email{apopov@hawaii.edu}).}}

\usepackage{amsopn}

\begin{document}

\pgfplotsset{clean/.style={axis lines*=left,
        axis on top=true,
        axis x line shift=0.0em,
        axis y line shift=0.75em,
        every tick/.style={black, thick},
        axis line style = ultra thick,
        tick align=outside,
        clip=false,
        major tick length=4pt}}

\maketitle

\begin{abstract}
AI and data-driven models have large potential for data assimilation applications by creating fast and accurate forecasts.
Their tendency to produce spurious inaccurate, nonphysical results---hallucination---however, raises a serious question about their long-term use, and can be categorized as untrustworthy methods.
Theory-driven methods on the other hand are slow, but are capable of staying physically realistic due to their mathematical underpinning, and can be categorized as trustworthy methods.
We argue that by making use of these methods in tandem, it is possible to build a relative measure of trust between the theory-driven and data-driven methods that results in a combined trustworthy methodology.
We argue, and then show, that the bandwidth scaling factors in the kernel density estimates can be used to represent our trust in the theory-driven and data-driven models.
We provide for ways in which these measures of trust can be adaptively computed through an expectation-maximization approach.
We combine all of these ideas to create the multifidelity ensemble Gaussian mixture filter and its adaptive trust version, which are particle filters capable of high-dimensional data assimilation.
We validate our ideas on both a static banana problem and on a sequential filtering example with the Lorenz '96 equations, showing that it is possible to create a particle filter that is capable of high dimensional convergent inference in the undersampled regime---when the number of theory-driven samples is less than the dimension of the system.
\end{abstract}

\begin{keywords}
trust, artificial intelligence, data assimilation, multifidelity inference, ensemble Gaussian mixture filter
\end{keywords}

\begin{MSCcodes}
60G25, 62L12, 62M20, 93E11
\end{MSCcodes}

\section{Introduction}
Data assimilation~\cite{reich2015probabilistic,asch2016data} combines information from a dynamical model prediction and independent sensor measurements in a probabilistic framework.
Multilevel Monte Carlo (MLMC)~\cite{Giles_2008_MLMC,Giles_2015_MLMC} combines different level of models to create better and cheaper estimators.
MLMC was extended to data assimilation with the EnKF~\cite{Hoel_2016_MLEnKF,Chernov_2017_MLEnKF,Kikuchi_2015_ROM-EnKF}, sampling ensemble smoothers~\cite{Sandu_2016_reduced-sampling4DVar}, and some particle filters~\cite{Reich_2016_MLETPF,Gregory_2017_MLETPF}.
Multifidelity data assimilation built on the MLMC idea to enable the use of more than one type of dynamical model for use in data assimilation applications.
Previous work introduced the multifidelity ensemble Kalman filter~\cite{popov2021multifidelity}, the multifidelity Kalman filter and multifidelity observations~\cite{popov2022multifidelityDA}, non-linear relations between models~\cite{popov2022multifidelity}, hyperreduced reduced order models ~\cite{donoghue2022multi}, adaptive hierarchical methods~\cite{silva2025adaptive}, and model forest generalizations~\cite{popov2022modelforest}.

While making significant gains in the past several years~\cite{karpatne2024knowledge,karpatne2017theory,cheng2023machine}, it is now well known that artificial intelligence-driven and other data-driven methods for scientific applications overstate both their speed and accuracy~\cite{mcgreivy2024weak}, often produce untrustworthy results~\cite{cheetham2024artificial,hicks2024chatgpt}, and have claims of superiority that have led to a reproducibility crisis~\cite{kapoor_leakage_2023}. 
This naturally gives rise to a problem: what steps can be taken to increase our trust~\cite{afroogh2024trust} in using such models?
Instead of increasing the reliability and trustworthiness of data-driven models themselves, this work aims to completely sidestep this issue through the use of multifidelity inference.
Multifidelity inference can couple models in a model hierarchy with one principal model towering over other, lesser models.
In this work we aim to show that if the principal model is a theory-driven model, and the lesser model is a data-driven model, e.g. one constructed using AI-like methods, then it is still possible to have overall trustworthy inference.
Previous work in particular~\cite{popov2022multifidelity} showed that even poor autoencoder-based reduced order models could aide in reducing error as long as they were coupled to a trustworthy full order model. 
While this work is targeting the problem of trust of artificial intelligence models as a particular class of data-driven models in data assimilation as that is of particular focus and importance at the time of this writing, the vast majority of this work is applicable to all manner of untrustworthy data-driven models, and not exclusive to artificial intelligence models.

Previous multifidelity work has made use of the (ensemble) linear control variate framework~\cite{popov2021multifidelity}, meaning that while coupling between model spaces could be non-linear, the coupling of information made use of an approximation to the best linear unbiased framework.
Effectively this means that the inference being performed was non-Bayesian~\cite{jaynes2003probability}, as all available information was not being utilized in the optimal fashion.
While linear coupling has proven to be powerful for the high dimensional setting, the low and medium dimension settings can take advantage of non-linear coupling through the use of convergent particle filters.
In particular, in this work, we focus on the ensemble Gaussian mixture filter (EnGMF)~\cite{popov2024adaptive,yun2022kernel,liu2016efficient,anderson1999monte} which makes use of kernel density estimation~\cite{silverman2018density} and the Gaussian mixture update to create samples that hopefully align with the true posterior distribution for a reduced number of samples compared to a bootstrap particle filter.
The EnGMF is used to perform optimal (in the limit of ensemble size) coupling between the information from the models, and is then also subsequently used to perform the coupling to sensor measurements.

We extend the philosophy behind the kernel density estimation used in the EnGMF to view the bandwidth scaling factor as a measure of trust (see~\cite{jsang2018subjective,shafer2020mathematical}) in the set of samples. We argue that large bandwidth scaling factors show a decrease either our trust in the samples themselves or the model forecast from which the samples derive. 
We then extend this idea to the multifidelity setting and argue that the coupling of bandwidth factors 
This means that a large bandwidth scaling factor on the samples derived from an data-driven model signals our mistrust of said model.
Using techniques developed in~\cite{popov2024adaptive} we then extend this notion to algorithmically determine our trust (through bandwidth scaling factor) in the data-driven model through an expectation maximization procedure.

The goals of this work are threefold. First, this work generalizes linear control variate coupling to non-linear couplings through the lens of Bayesian inference coupling information from trustworthy theory-driven models and untrustworthy data-driven models, and builds in a method of trust in each of the models through bandwidth scaling factors in kernel density estimation. Second, this work introduces the the multifidelity ensemble Gaussian mixture filter (MFEnGMF), which we show is a convergent particle filter for the undersampled regime---when the number of particles is less than the size of the state space for systems whose dimension is larger than is feasible to estimate with traditional particle filters.
Third, we show how our proposed measures of trust can be adaptively determined in time through an expectation-maximization procedure.

This paper is organized as follows:
we begin with some background on data assimilation, the ensemble Gaussian mixture filter, the multifidelity ensemble Kalman filter, reduced order modeling, and expectation maximization parameter estimation in~\cref{sec:background}.
We introduce the multifidelity ensemble Gaussian mixture filter, our trust framework that is embedded therein, and an adaptive trust algorithm in~\cref{sec:mfengmf}. 
Penultimately in \cref{eq:numerical-experiments} this work validates the proposed methodology and theory through practical numerical experiments on a two-dimensional static `banana' problem and on the Lorenz '96 equations.
Finally we conclude with some closing remarks in~\cref{sec:conclusions}.

\section{Background}
\label{sec:background}

Assume that there is some underlying true state $\kappa$ about which we have some uncertainty represented by $p_X$ which is the probability distribution of the random variable $X$, which we call the principal variate, being the prior source of information.
In this work the principal variate represents the information obtained from the theory-driven model, thus is henceforth referred to as the theory-driven variate.
We can represent this relationship as,
\begin{equation}
    X = \kappa + \eta_X,\quad
\end{equation}
where $\eta_X$ is additive error whose distribution represents our underlying uncertainty about the deviation of our information $X$ from the truth $\kappa$.
Assume additionally that we have uncertainty about some transformation of the truth into a (usually lower dimensional) space,
\begin{equation}\label{eq:lambda}
    \upsilon = \lambda(\kappa) + \eta_\upsilon,
\end{equation}
where $\upsilon$ is a random variable whose distribution $p_\upsilon$ represents our uncertainty about $\kappa$ transformed by the deterministic function $\lambda$ with error $\eta_\upsilon$ similar to the above. We call the variable $\upsilon$ the ancillary variate, being the new source of information. 
In this work we make use of the above formulation in coupling random variables for two distinct purposes, meaning that the operator $\lambda$ in~\cref{eq:lambda} can stand for multiple different things.
One use of $\upsilon$ is that of the observations---real world data about which we have uncertainty. In this case we denote the random variable by $Y$ and the transformation as the (non-)linear observation operator $h$. 
The other use of $\upsilon$ is to denote our uncertainty coming from a surrogate model, which for the purposes of this work we assume to be an untrustworthy data-driven model. In this case we denote the random variable as $U$ and call it the data-driven variate and the transformation as the (non-)linear encoder $\theta$. 

\subsection{Reduced order and multifidelity modeling}

The first purpose of the formulation is to perform reduced order modeling. 
Take $X$ to be the random variable representing our information about the state of interest.
Our goal is to define a lower dimensional encoding of most of the information contained in $X$, denoted by $U$,  that can be propagated in time through an data-driven model.
We defined the dimensionality reduction of $X$ into the space of $U$ by,
\begin{equation}\label{eq:order-reduction}
    U = \theta(X) + \eta_U,
\end{equation}
where the operator $\theta$ corresponds to one of the forms of $\lambda$ in~\cref{eq:lambda}, and the 

Correspondingly, the reverse operation,
\begin{equation}
    X \approx \phi(\theta(X)),
\end{equation}
uplifts the reduced order variable back into the full space, and ideally retains as much information about $X$ as possible.

For our theory-driven variate we take the random variable $X$ to represent some information about the state of our system of interest.
This theory-driven variate represents our uncertainty about the state, and is therefore trustworthy information about the system. 

\subsection{Data Assimilation and the Ensemble Kalman Filter}

Explicitly in the context of data assimilation, we are given prior information in the form of a theory-driven variate, $X^-$, where the minus means ``prior'', and we are additionally given measurement information in the form,
\begin{equation}
    y = h(\kappa) + \eta_Y,
\end{equation}
where $h$ is a measurement function corresponding to one of the forms of $\lambda$ in~\cref{eq:lambda}, and for this work $\eta_Y$ is considered to be unbiased additive error with covariance of $R$. Data assimilation is concerned with finding a description of the distribution $p_{X^- | Y = y}$ in a computationally efficient manner.

The ensemble Kalman filter (EnKF)~\cite{evensen1994sequential, burgers1998analysis} takes advantage of the fact that ensemble propagation is extremely effective at capturing uncertainty about highly non-linear dynamical systems~\cite{kalnay2003atmospheric}. As such, our uncertainty about $X^-$ is expressed as a collection of points---which are not, but can be seen as samples---that are computationally useful descriptors about which sample statistics can be computed,
\begin{equation}
    E_{X^-} = [x^-_1, x^-_2, \dots, x^-_{N_X}],
\end{equation}
which is commonly referred to as an ensemble of size $N_X$.

The perturbed observations ensemble Kalman filter~\cite{evensen1994sequential, burgers1998analysis} create an approximation to the Kalman filter equations~\cite{kalman1960new} through statistical covariances and ensembles,
\begin{equation}\label{eq:EnKF}
    E_{X^+} = E_{X^-} - \widetilde{K}(h(E_{X^-}) - E_{Y}),
\end{equation}
where $\widetilde{K}$ is the stochastic Kalman gain derived from the ensemble $E_{X^-_k}$, and $E_{Y}$ is the ensemble of `perturbed observations' after which this version of the algorithm is named, discussed later.
We note that the stochastic Kalman gain is a biased estimator of the Kalman gain~\cite{popov2020explicit}.
In this work we make use of the linearized~\cite{michaelson2023ensemble} form of the Kalman gain that is applied to every single ensemble member individually, meaning that the Kalman gain for the ensemble member with index $i$  is,
\begin{equation}\label{eq:EnKF-statistical-Gain}
    \widetilde{K} = \widetilde{\Sigma}^{X^-}H^T_i\left( H_i\widetilde{\Sigma}^{X^-}H^T_i + R\right)^{-1},
\end{equation}
where $\widetilde{\Sigma}^{X^-}$ is the statistical covariance of $X^-$ and $H_i$ is the Jacobian of $h$ computed at ensemble member $x^-_i$.
It is important to note that the ensemble $E_{X^-}$ in~\cref{eq:EnKF} has to first be inflated, thus every $i$th element of the ensemble is first modified,
\begin{equation}
    x^-_1 \leftarrow \widetilde{\mu}^{X^-} + \alpha_X\left(x^-_1 - \widetilde{\mu}^{X^-}\right),
\end{equation}
where $\widetilde{\mu}^{X^-}$ is the statistical mean, and where the scaling inflation factor $\alpha_X \geq 1$ attempts to compensate for the fact that the stochastic Kalman gain $\widetilde{K}$ is a biased estimator~\cite{popov2020explicit}.

The ensemble of `perturbed observations', $E_{Y_k}$ has to have the following properties:
\begin{equation}\label{eq:perturbed-observations}
    \mathbb{E}{E_{Y_k}} = y, \quad \mathbb{V}{E_{Y_k}} = R,
\end{equation}
meaning that the mean of the ensemble is the observation and the covariance is the measurement covariance.
Typically, this ensemble of perturbed observations is constructed by sampling each member from the Gaussian distribution with mean $y$ and covariance $R$. This is what is used in this work, but we note that any distribution with those two properties could be used.

\subsection{Multifidelity ensemble Kalman filter and Linear Coupling}

In the context of multifidelity inference, we now combine the two concepts of order reduction and data assimilation in order to combine information both from multiple full order and reduced order models and from observations.
The MFEnKF as originally posed in~\cite{popov2021multifidelity} is a sequential filter and is not suitable for single-step filtering. 
In this work we present a simplification and modification to the MFEnKF that is capable of making the MFEnKF a single-step filter.

\subsubsection{Multifidelity ensemble propagation}

In the propagation step, we need to propagate the theory-driven and data-driven ensembles through the theory-driven and data-driven models with the random variables that represents our uncertainty in said models, sometimes called model error or process noise.
Denote the propagation of the theory-driven ensemble through the theory-driven model by,
\begin{equation}
    E_{X_k} = \mathcal{M}_X(E_{X_{k-1}}, \xi_{X_{k-1}}),
\end{equation}
where $\xi_{X_{k-1}}$ is the random variable whose distribution represents our uncertainty about the theory-driven model propagation with respect to the truth.
Denote the propagation of the data-driven ensemble through the data-driven model by,
\begin{equation}
    E_{U_k} = \mathcal{M}_U(E_{U_{k-1}}, \xi_{U_{k-1}}),
\end{equation}
where $\xi_{U_{k-1}}$ is the random variable whose distribution represents our uncertainty about the theory-driven model propagation with respect to the truth.

\subsubsection{Multifidelity Ensemble Kalman Filter update}

In the original~\cite{popov2021multifidelity} and subsequent followup works~\cite{popov2022multifidelity,popov2022modelforest}, the MFEnKF has been posed as acting on propagated control variate information. In this work we modify the MFEnKF to function in a linearized~\cite{michaelson2023ensemble} and single-step filtering fashion.
All the following derivations are assumed to occur at time index $k$, and thus the time index will be omitted.
Suppose that we have a dimensionality reduction technique 

The most important part of the MFEnKF is the linear coupling that is enforced between the theory-driven variate and the control and data-driven variates.
\begin{equation}\label{eq:linear-coupling}
    Z^- = X^- - S(\hat{U}^- + U^-),
\end{equation}
where the matrix $S$ is a gain matrix that represents the optimal coupling. 
In fact, the optimal coupling is the Kalman gain,
\begin{equation}\label{eq:optimal-gain}
    S = \Sigma_{X^-, \hat{U}^-}{\left(\Sigma_{\hat{U}^-, \hat{U}^-} + \Sigma_{{U}^-, {U}^-}\right)}^{-1},
\end{equation}
where, under certain simplifying assumptions (see~\cite{popov2021multifidelity}), it can be shown that,
\begin{equation}
    S = \frac{1}{2}\Phi,
\end{equation}
where $\Phi$ is the Jacobian of $\phi$ in~\cref{eq:order-reduction}.

The multifidelity ensemble Kalman filter proceeds as follows:
\begin{equation}\label{eq:MFEnKF}
    \begin{aligned}
        E_{X^+} &= E_{X^-} - \widetilde{K}(h(E_{X^-}) - E_{Y}^{X}),\\
        E_{\hat{U}^+} &= E_{\hat{U}^-} - \Theta \widetilde{K}(h(\phi(E_{\hat{U}^-})) - E_{Y}^{X}),\\
        E_{U^+} &= E_{U^-} - \Theta \widetilde{K}(h(\phi(E_{U^-})) - E_{Y}^{U}),\\
    \end{aligned}
\end{equation}
where $\Theta$ is the linearization of~\cref{eq:order-reduction}.
Instead of~\cref{eq:EnKF-statistical-Gain}, the statistical Kalman gain is now,
\begin{equation}
    \widetilde{K} = \widetilde{\Sigma}^{Z^-}H^T_i\left( H_i\widetilde{\Sigma}^{Z^-}H^T_i + R\right)^{-1},
\end{equation}
where the covariance $\widetilde{\Sigma}^{Z^-}$ is computed through the constituent ensembles (see~\cite{popov2021multifidelity}).
The perturbed observations have the same properties as in~\cref{eq:perturbed-observations}, with the perturbations $E_{Y}^{U}$ being fully independent of the perturbations $E_{Y}^{X}$. Unlike~\cite{popov2021multifidelity}, in this work the perturbed observations share the same covariance. Additionally no ``mean correction'' heuristic is applied.
As with the EnKF, the linearized version of all the updates is used in this work.

\subsection{Ensemble Gaussian Mixture Filter}

The ensemble Gaussian mixture filter takes as input samples that represent our uncertainty about some prior distribution and outputs

Instead of making the simplifying assumption that our samples are the only information that we have available---as is made by particle filters like the bootstrap particle filter---we instead make the assumption that our samples come from some unknown underlying distribution.

\subsubsection{Kernel density estimation}

Take an arbitrary ensemble $E_X$ of samples from a probability distribution $p_{X}$. From these samples, it is possible to construct a Gaussian mixture kernel density estimate~\cite{silverman2018density},
\begin{equation}\label{eq:KDE}
    \widetilde{p}_X(x) = \sum_{i=1}^{N_X} w_i^X \mathcal{N}(x ; \mu^X_i, \Sigma^X_i),
\end{equation}
where the weights $w_i^X$ correspond to the weights of the samples (typically taken to be $w_i^X = 1/N_X$ in the case of equally weighted samples, the means are taken to be the samples themselves, meaning that,
\begin{equation}
    E_X = [\mu^X_1, \mu^X_2, \dots, \mu^X_{N_X}],
\end{equation}
and the covariance is taken to be,
\begin{equation}\label{eq:covariance-KDE}
     \Sigma^X_i = (s_X h_X)^2 \operatorname{Cov}(E_X),
\end{equation}
where $\operatorname{Cov}(E_X)$ is the ensemble covariance, $h_X$ is the optimal, Silverman~\cite{silverman2018density} bandwidth defined as,
\begin{equation}\label{eq:Silverman}
    h_X = \left(\frac{4}{N_X(n + 2)}\right)^{\frac{1}{n + 4}},
\end{equation}
in the case that the underlying distribution has Gaussian curvature, and $s_X$ is a scaling factor that accounts for the discrepancy between the true distribution from which the samples are taken, and the Gaussian assumption underpinning the estimate of the Silverman bandwidth~\cite{popov2024Epanechnikov}.

\subsubsection{The scaling factor as a measure of trust}
\label{sec:scaling-factor-discussion}

The scaling factor can play multiple roles in determining the utility of the KDE defined above. 
As mentioned before $h_X$ in~\cref{eq:Silverman} is optimal provided that the underlying distribution is Gaussian.
This means that the scaling factor $s_X$ can account for the non-Gaussian curvature of the real underlying distribution of the data, as the optimal bandwidth is often not the Silverman bandwidth~\cite{popov2024adaptive}. 

The second role of the scaling factor is in determining the conservativeness of the kernel density estimate.
It can be shown~\cite{liu2016efficient} that if the covariance in~\cref{eq:covariance-KDE} is used for the kernel density estimate in~\cref{eq:KDE}, that the covariance of the distribution
defined is (with heavy abuse of notation),
\begin{equation}
    \operatorname{Cov}\left(\widetilde{p}_X\right) = \left(1 + (s_X h_X)^2\right) \operatorname{Cov}(p_X),
\end{equation}
meaning that the covariance of the KDE is a conservative estimate of the true underlying covariance. 
This means that as the scaling factor decreases, the covariance becomes less and less conservative.

The third role of the scaling factor, and the one with which this paper is concerned, is in the trust that is placed in the samples. 
The optimal bandwidth~\cref{eq:Silverman} is derived to minimize the (asymptotic) mean integral squared error between the KDE estimate in~\cref{eq:KDE} and the true underlying distribution of the samples.

There are two very important caveats to this: first, the KDE is only optimal in the sense of mean integral squared error, and it is not necessarily the case in any one particular instance of an ensemble.
This means that one collection of samples could induce a KDE that is a poor representation of the underlying distribution.
This is particularly of note when the ensemble size $N_X$ is small and the samples are almost always provide a poor representation of the underlying distribution.
In this regime, an increase  scaling factor $s_X$ can counter the fact that our collection of samples is likely not representative of what we are actually trying to approximate.

The second caveat is pertinent to multifidelity inference: the ensemble of samples that we are given actually has to come from the distribution that are attempting to estimate.
In other words, if we have a collection of samples from a untrustworthy data-driven model, then there is always a large mismatch between the distribution that we can approximate by KDE and the distribution that we actually care about, no matter the ensemble size $N_X$.
If we are reasonably confident that the ensemble comes from a distribution that is `nearby' to the distribution that we actually care about, then a scaling factor $s_X$ on the order of unity is reasonable.
If on the other hand this is not the case, then a larger scaling factor would express our mistrust in model from which the ensemble originates.

\subsubsection{Gaussian mixture update}

We focus on the case when $X$ and $\upsilon$ are independent and whose distributions are represented by Gaussian mixtures,
\begin{equation}
    p_X(x) = \sum_{i=1}^{N_X} w_i^X \mathcal{N}(x ; \mu^X_i, \Sigma^X_i),
\end{equation}
for $X$, where $N_X$ is the number of components, $w_i^X$ are weights that sum to one, $\mu^X_i$ are the means of each component, and $\Sigma^X_i$ are the covariances of each component, and,
\begin{equation}
     p_\upsilon(u) = \sum_{j=1}^{N_\upsilon} w_j^\upsilon \mathcal{N}(u ; \mu^\upsilon_j, \Sigma^\upsilon_j),
\end{equation}
for $\upsilon$, where $N_\upsilon$ is the number of components, $w_j^\upsilon$ are weights that sum to one, $\mu^\upsilon_j$ are the means of each component, and $\Sigma^\upsilon_j$ are the covariances of each component.

Our goal is to build a representation of the total information given by the distribution of $\zeta = X | \upsilon$.
This can be approximated through the Gaussian mixture,
\begin{equation}\label{eq:posterior-GMM}
    p_\zeta(z) =  \sum_{i=1}^{N_X}\sum_{j=1}^{N_\upsilon} w^\zeta_{i,j} \mathcal{N}(z ; \mu^\zeta_{i,j}, \Sigma^\zeta_{i,j}),
\end{equation}
by making use of the Gaussian mixture update, for $i = 1, \dots, N_X$, and for $j = 1, \dots, N_\upsilon$,
\begin{equation}\label{eq:GMU}
\begin{aligned}
    \mu^\zeta_{i,j} &= \mu^X_i - G_{i,j}(\lambda(\mu^X_i) - \mu_j^\upsilon),\\
    \Sigma^\zeta_{i,j} &= (I - G_{i,j}\Lambda_i)\Sigma^X_i,\\
    w^\zeta_{i,j} &\propto \mathcal{N}\left(\lambda(\mu^X_i)\,;\, \mu_j^\upsilon, \Lambda_i \Sigma_i^X\Lambda_i + \Sigma_j^\upsilon\right)\\
    G_{i,j} &= \Sigma^X_i \Lambda_i^T{\left(\Lambda_i\Sigma^X_i \Lambda_i^T + \Sigma^\upsilon_j\right)}^{-1},\\
    \Lambda_i &= \lambda'(\mu^X_i),  
\end{aligned} 
\end{equation}
where $\mu^\zeta_{i,j}$ are the means of \cref{eq:posterior-GMM}, $\Sigma^\zeta_{i,j}$ are the covariances, $w^\zeta_{i,j}$ are the weights (summing to unity), $G_{i,j}$ are gain matrices (despite similar structure, these are not the Kalman gain, and are unique per mode), and $\Lambda_i$ are the Jacobians of the measurement function.
This can be though of as performing $N_X N_\upsilon$ distinct extended Kalman filter updates~\cite{hanebeck2025ensemble}.

Typically to avoid numerical degeneracy, the weights are modified by a defensive factor,
\begin{equation}
     w^\zeta_{i,j} \leftarrow  (1 - \delta)w^\zeta_{i,j} + \delta (N_X N_\upsilon)^{-1},
\end{equation}
where $\delta$ for this work is taken to be $\delta = 10^{-4}$. If $\delta$ is zero then the weights are kept the same, and if the defensive factor is one then the weights are all uniform, meaning that the above is a continuous interpolation between the EnGMF weights and uniform weights. This means that the defensive factor acts as a regularizer.

For future compactness of notation we can write the Gaussian mixture of means $\mu$, covariances $\Sigma$, and weights $w$ as,
\begin{equation}
    \mathcal{GM}(\mu, \Sigma, w),
\end{equation}
meaning that the result of the Gaussian mixture update in~\cref{eq:GMU} can be written as,
\begin{equation}
    \mathcal{GM}(\mu^\zeta, \Sigma^\zeta, w^\zeta) = \operatorname{GMU}_\lambda\left(\mathcal{GM}(\mu^X, \Sigma^X, w^X), \mathcal{GM}(\mu^\upsilon, \Sigma^\upsilon, w^\upsilon)\right),
\end{equation}
where the action of the Gaussian mixture update, dependent on $\lambda$, is denoted by $\operatorname{GMU}_\lambda$.

\subsubsection{Sampling from a Gaussian mixture}

As the propagation step relies on a collection of samples, and the Gaussian mixture update produces an estimate of the posterior distribution, in order to be able to propagate samples, it becomes imperative to derive samples therefrom. In this section we describe a simple method that allows us to create samples from said Gaussian mixture posterior.
Given a Gaussian mixture $\mathcal{GM}(\mu, \Sigma, w)$, a sample $x$ is generated as follows:
\begin{enumerate}
    \item Take the discrete distribution induced by the weights $w$ and sample a component index $\ell$. This can be performed efficiently through some discrete sampling algorithm.
    \item Sample $x$ from the normal distribution $\mathcal{N}(\mu_\ell, \Sigma_\ell)$.
\end{enumerate}
This procedure can be repeated as many times as necessary to the desired number of samples.

Of particular note is that the number of samples required from the distribution can be significantly smaller than the number of components of the Gaussian mixture model. In that particular case, for this work the divide and conquer algorithm described in~\cite{popov2026divideconquerstrategymultinomial} is utilized to significantly speed up the computation.

\subsubsection{EnGMF Algorithm}
We now combine all the previous elements, including kernel density estimation, the Gaussian mixture update, and the resampling procedure in order to describe the EnGMF algorithm.
The ensemble Gaussian mixture filter proceeds as follows:
\begin{enumerate}
    \item Take the prior ensemble $E_{X^-}$ and use kernel density estimation with bandwidth $h_X$ and scaling factor $s_X$ to create a Gaussian mixture distribution that represent an approximation to our uncertainty about the prior information,
    \begin{equation}
        p(X^-) = \mathcal{GM}(\mu^{X^-}, \Sigma^{X^-}, w^{X^-}),
    \end{equation}
    where we denote our approximation of our prior uncertainty through the distribution $p(X^-)$.
    It is important to note that if we were to use the standard kernel density estimate of the covariance, each covariance matrix in $\Sigma^{X^-}$ would be parameterized by the bandwidth scaling factor $s_X$, as we aim to use said bandwidth scaling factor as a measure of trust in the prior ensemble and model from which it came.
    \item We update the distribution by making use of observations,
    \begin{equation}
    \begin{aligned}
        p(X^+) &= p(X^- | Y) = \operatorname{GMU}_{h}(p(X^-), p(Y)),\\
        &= \mathcal{GM}(\mu^{X^+}, \Sigma^{X^+}, w^{X^+}),
    \end{aligned}
    \end{equation}
    \item finally we resample $N_X$ samples $E_{X^+}$ from $p(Z^+)$.
\end{enumerate}

\noindent The samples can then be propagated in time and the algorithm run over and over for sequential data assimilation.

\subsubsection{Adaptive Trust Determination}
\label{sec:adaptive-trust}
As discussed previously in~\cref{sec:scaling-factor-discussion}, the bandwidth scaling factor, $s_X$, can be viewed as a measure of trust in the prior ensemble.
If we are able to determine the bandwidth scaling in an adaptive fashion this would be an analogy for determining the trust that we are putting 
In~\cite{popov2024adaptive} an algorithm for determining the bandwidth factor $h_X^-$ in an adaptive fashion was proposed.
This section outlines a modified expectation maximization~\cite{neal1998view,bocquet2020bayesian, bishop2006pattern} algorithm for determining the bandwidth scaling factor $s_X$. 
The expectation maximization that determines the best scaling factor runs in two steps:
\begin{enumerate}
    \item \textit{Expectation}: for the $i$th step of the algorithm, construct an objective function in the joint density of the prior, measurement, and bandwidth scaling factor, 
    \begin{equation}
        J(s_X; s_X^{(i)}) = \mathbb{E}_{x \sim p\left(X^+ \middle| s_X^{(i)}\right)} \log \left[p(x; Y | X^-)p(x; X^- | s_X)\right],
    \end{equation}
    where the expectation is over samples from the posterior with the  previous value of the scaling factor $s_X^{(i)}$.
    \item \textit{Maximization}: for the $i$th step of the algorithm, the objective function is maximized to find the next value of the bandwidth,
    \begin{equation}
        s_X^{(i+1)} = \operatorname{arg\,max}_{s_X} J(s_X; s_X^{(i)}),
    \end{equation}
    where the the expectation in the cost is calculated through sampling.
\end{enumerate}
For a practical implementation for this work, we initially take a naive guess of the bandwidth scaling factor $s_X^{(0)}=1$ as the nominal bandwidth scaling.
For sequential filtering, the initial bandwidth scaling factor is taken to be the value determined at the previous step, as we assume that our trust in the theory-guided ensemble does not change much over time.
In the interest of fast computation, only one full step of expectation maximization is performed, with the cost function approximated by $N_X$ samples at each step, and the maximization performed through a few steps of gradient ascent with a fixed step size. For this work the gradient of the cost function is computed using a forward finite difference method in the interest of ease of implementation.

The EnGMF with adaptive trust determination through the bandwidth scaling factor we term the AEnGMF.

\section{The Multifidelity Ensemble Gaussian Mixture Filter}
\label{sec:mfengmf}

\begin{figure}
    \centering
    \includegraphics[width=0.99\linewidth]{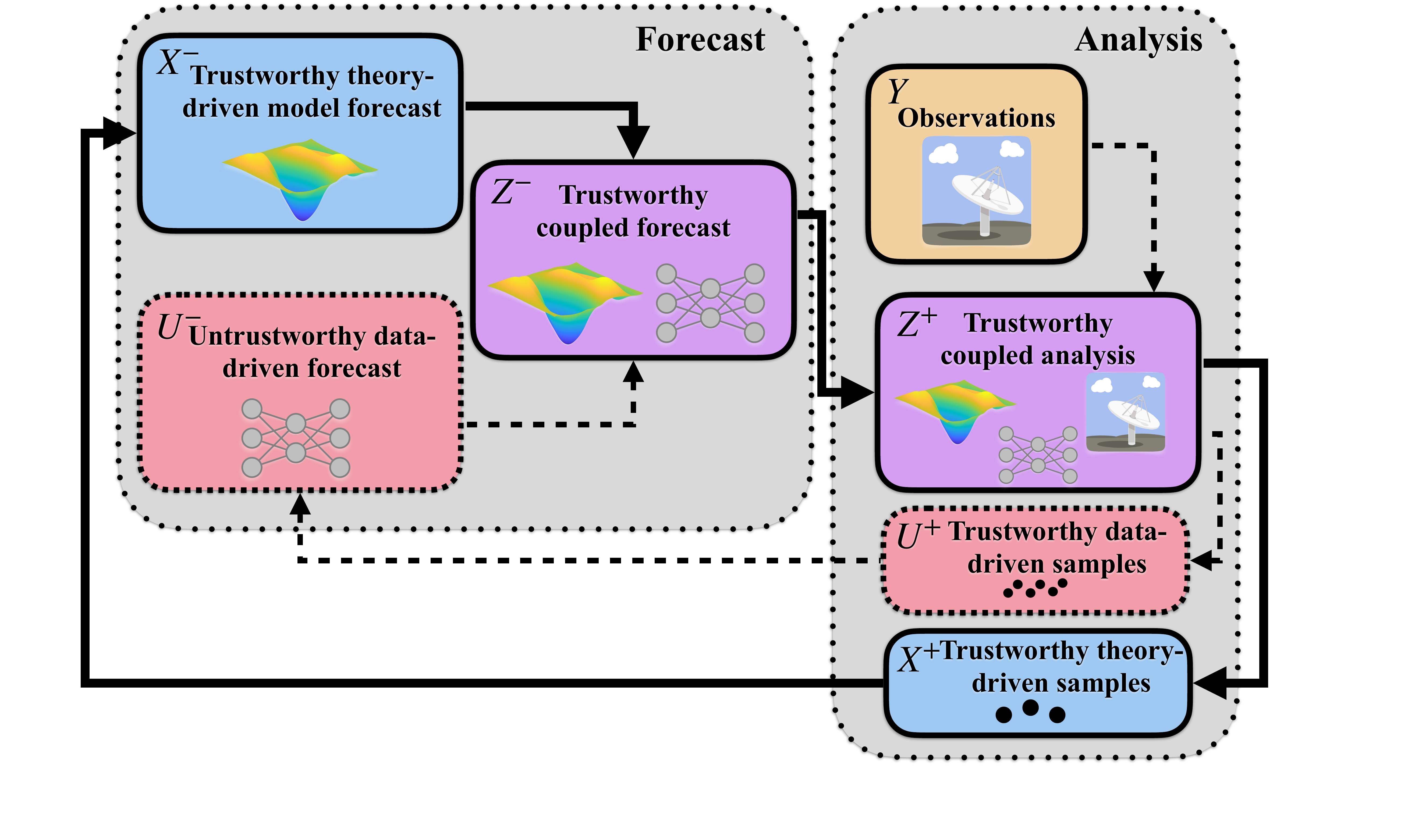}
    \caption{A description of the generalized multifidelity ensemble framework. The forecast step consists of the trustworthy theory-driven model update, untrustworthy data-driven model update, and the coupling of the two forecasts to create a trustworthy coupled forecast. The analysis step then fuses in external observations to create the trustworthy coupled analysis, from which trustworthy samples are created for both the theory-driven and data-driven models. The cycle then repeats ad infinitum.}
    \label{fig:MFEnGMF}
\end{figure}

The key observation that allows us to generalize the multifidelity ensemble Kalman filter to non-linear couplings and non-linear filtering is the following:  
in the case where all our uncertainty is Gaussian, and all our functions are linear, then the coupling in~\cref{eq:linear-coupling} defines a conditioning of the theory-driven variate on the data-driven variate:
\begin{equation}\label{eq:conditional}
    p(Z^-) = p(X^- | U^-),
\end{equation}
meaning that a non-linear, more exact coupling would take the form of a better approximation of the exact conditional.
Building on the ensemble Gaussian mixture filter, we propose to approximate the distribution of $X^-$ and $U^-$ by Gaussian mixture models, and approximating the conditional distribution in~\cref{eq:conditional} through a Gaussian mixture update. The rest of the algorithm follows the typical EnGMF path. 
This non-linear coupling between the theory-driven and data-driven models through this Gaussian mixture update creates a filter that we term the multifidelity ensemble Gaussian mixture filter, or MFEnGMF for short.

The multifidelity Gaussian mixture filter proceeds in the following broad steps: first, the forecast (prior) ensembles $E_{X^-}$, and $E_{U^-}$ are converted into Gaussian mixture model approximations of their underlying distributions. A Gaussian mixture update is performed to beget a Gaussian mixture approximation to the prior total variate $Z^-$. 
An observation $y$, and corresponding Gaussian mixture error is obtained, and another Gaussian mixture update is performed to obtain a Gaussian mixture approximation of the analysis (posterior) $Z^+$. A resampling procedure is performed to get samples from $Z^+$. Part of the samples are converted into the space of the reduced order model and are treated as samples from $U^+$ while the rest are treated as samples from $X^+$. The samples are propagated in time and the process repeats ad infinitum.
The framework is loosely visually described in~\cref{fig:MFEnGMF}. 
A detailed algorithmic description of the update follows:
\begin{enumerate}
    \item Take the prior ensembles $E_{X^-}$, and $E_{U^-}$ and use kernel density estimation to create Gaussian mixture distributions that represent approximations to our uncertainty about the prior information,
    \begin{equation}\label{eq:pXminus-pUminus}
    \begin{aligned}
        p(X^-) & = \mathcal{GM}(\mu^{X^-}, \Sigma^{X^-}, w^{X^-}),\\
        p(U^-) &= \mathcal{GM}(\mu^{U^-}, \Sigma^{U^-}, w^{U^-}),
    \end{aligned}
    \end{equation}
    where we denote our approximation of our theory-driven prior uncertainty through the distribution $p(X^-)$ and our approximation of our data-driven prior uncertainty as $p(U^-)$.
    It is important to note that when the standard kernel density estimation technique is used, the covariances $\Sigma^{X^-}$ and $\Sigma^{U^-}$ are parameterized by the bandwidth scaling factors $s_X$ and $s_U$ respectively, as these values are used as a measure of trust in each of the models.
    \item Through the use of the control coupling, we build a prior total variate distribution:
    meaning that we build the Gaussian mixture,
    \begin{equation}\label{eq:pZminus}
    \begin{aligned}
        p(Z^-) &= p(X^- | U^-) = \operatorname{GMU}_{\theta}(p(X^-), p(U^-)),\\
        &= \mathcal{GM}(\mu^{Z^-_k}, \Sigma^{Z^-_k}, w^{Z^-_k}),
    \end{aligned}
    \end{equation}
    through the use of the Gaussian mixture update using the encoder $\theta$.
    \item next we update the distribution by making use of observations,
    \begin{equation}
    \begin{aligned}
        p(Z^+) &= p(Z^- | Y) = \operatorname{GMU}_{h}(p(Z^-), p(Y)),\\
        &= \mathcal{GM}(\mu^{Z^+_k}, \Sigma^{Z^+_k}, w^{Z^+_k}),
    \end{aligned}
    \end{equation}
    through the use of the Gaussian mixture update using the observation $y$ and the observation operator $h$, and
    \item finally we resample, using an algorithm such as~\cite{popov2026divideconquerstrategymultinomial}, $N_X + N_U$ samples from $p(Z^+)$, of which  $N_X$ samples we take as $E_{X^+}$ and $N_U$ samples we transform using the encoder $\theta$ to, $E_{U^+}$.
\end{enumerate}

\noindent We first discuss the issue of resampling in the MFEnGMF. Instead of the MFEnKF approach~\cref{eq:MFEnKF} where the three variates are split into three distinct Kalman updates, the MFEnGMF creates samples that resemble the total variate posterior $Z^+$. For any one given step this is fine given the trivial observation that $Z^+ = X^+ | U^+$. 
However in the case of a Gaussian mixture update, this observation is violated.
This is largely mitigated again under the assumption that the ensembles $E_{X^+}$ and $E_{U^+}$ largely portray different information under model propagation at the next time step.

We next discuss the convergence, in ensemble size, of the method above. When the data-driven scaling factor $s_U$ is sufficiently large, the Gaussian mixture update conditioning on the data-driven variate $U^-$ tends towards the variate $X^-$, meaning that the method converges to that of the EnGMF.
In the case that $s_U$ is not sufficiently large, the MFEnGMF does something different. 
In the Bayesian inference sense, as the MFEnGMF takes advantage of additional information through the data-driven model, the MFEnGMF is---in the very loose sense---more Bayesian than the EnGMF, but can produce a worse estimate if our trust in the data-driven model is too high. Thus, it is paramount that we explore our trust in both the theory-driven ensemble and the data-driven model to create the best possible estimate of the true posterior.

\begin{remark}[Pruning, Merging, and Capping]
    The MFEnGMF suffers from a combinatorial explosion of Gaussian mixture terms. 
    For the MFEnGMF, the pruning, merging, and capping~\cite{durant2025kernel} call all help alleviate the need to compute a combinatorial explosion of Gaussian mixture terms, though this obvious direction is not explored in this work.
\end{remark}

\subsection{Trusting the theory-driven ensemble and data-driven model}
\label{sec:trust}

Previously in~\cref{sec:scaling-factor-discussion} we discussed the scaling factor $s_X$ as determining how trustworthy the kernel density estimate portion of the EnGMF is.
This section aims to expand those ideas to have a measure of trust not only for the theory-driven ensemble through the scaling factor $s_X$, but also for the data-driven model through the scaling factor $s_U$.

The scaling factors $s_X$ and $s_U$ determine the magnitude of the covariances in~\cref{eq:pXminus-pUminus} and have a significant impact on the distribution of the prior total variate in~\cref{eq:pZminus}. 
We term $s_X$ the theory-driven scaling factor, and $s_U$ the data-driven scaling factor.
As mentioned in~\cref{sec:scaling-factor-discussion}, we make use of these scaling factors in order to determine our relative trust in the theory-driven and data-driven models respectively.
We analyze the case where we have a surplus of samples from the untrustworthy data-driven model but are deficient in samples from the trustworthy theory-driven model.
This means that $N_U$ is large and that the probability distribution approximated by the kernel density estimate in the base case of the scaling factor $s_U=1$ is a good approximation of the uncertainty predicted by the data-driven model.
This also means that $N_X$ is small and that the probability distribution approximated by the kernel density estimate in the base case of the scaling factor $s_X=1$ is a poor approximation of the uncertainty predicted by the theory-driven model.

In the framework above, we have a small amount of good trustworthy samples from the theory-driven model that simply cannot adequately describe our uncertainty, and thus desperately need to be augmented with whatever information that is available, namely the untrustworthy samples from the data-driven model. We now look at several cases of modifying $s_X$ and $s_U$ and attempt to understand what happens to the distribution of $p(Z^-)$ as these scaling factors are modified.

We first look at the case that the data-driven scaling factor goes to infinity, meaning that $s_U\to\infty$.
This type of scaling means that each $\Sigma^{U^-}_j$ goes to infinity. 
Intuitively this means that the distribution 
Na\"ively in the limit of data-driven scaling factor $s_U$ going to infinity,
\begin{equation}
    p(X^-) = \lim_{s_U\to\infty} p(Z^-),
\end{equation}
meaning that \textit{increasing the data-driven scaling factor $s_U$ decreases our trust in the data-driven model}.

We then look at the case that the data-driven scaling factor goes to zero, meaning that $s_U\to0$. 
This type of scaling means that each $\Sigma^{U^-}_j$ goes to zero. 
In this case, the data-driven model ensemble means entirely replace the linear parts of the theory-driven model ensemble means, and the covariances degenerate to only contain uncertainty in the null space of the linearization of $\theta$, where that uncertainty is described by the covariances of $X^-$. 
This effectively means that \textit{decreasing the data-driven scaling factor $s_U$ increases the trust in the data-driven model, while simultaneously trusting the theory-driven model ensemble enough to describe the remaining uncertainty}.

We then look at the case that the theory-driven scaling factor goes to infinity, meaning that $s_X\to\infty$.
This type of scaling means that each $\Sigma^{X^-}_j$ goes to infinity. 
Intuitively this means that the theory-driven distribution becomes degenerate, and we are actually in a similar situation as the case where $s_U\to0$, where the data-driven model ensemble means are ignored, and the covariances degenerate to only contain uncertainty in the null space of the linearization of $\theta$.
This effectively means that  \textit{increasing the theory-driven scaling factor $s_X$ weakens our trust in the theory-driven model ensemble, while simultaneously decreasing trust in the data-driven model}.

We then look at the case that the theory-driven scaling factor goes to zero, meaning that $s_X\to0$.
This type of scaling means that each $\Sigma^{X^-}_j$ goes to zero. 
Again this means that the theory-driven distribution becomes degenerate, the data-driven model ensemble means are ignored, and the density of $Z^-$ closely matches that of the theory-driven model ensemble, but weighted completely by the data-driven model KDE.
This effectively means that \textit{decreasing the theory-driven scaling factor $s_X$ increases our trust in the theory-driven model ensemble, while simultaneously increasing trust in the data-driven model}.

We combine the observations above into the following rough guideline:

\begin{center}
\begin{tabularx}{\linewidth}{rXX}
     & \textbf{Increase} $s_X$ & \textbf{Decrease} $s_X$ \\
    \textbf{Increase} $s_U$ & Low trust in theory-driven model ensemble and slightly mistrust data-driven model.  & Heavily trust theory-driven model ensemble and slightly mistrust data-driven model.\\
    \textbf{Decrease} $s_U$ & Low trust in theory-driven model ensemble and highly trust data-driven model.  & High confidence in theory-driven model ensemble and highly trust data-driven model.
\end{tabularx}
\end{center}

\noindent Where, out of the four possibilities, for our target use case, increasing both $s_X$ and $s_U$ makes the most sense, as we don't want to trust our theory-driven model ensemble to be representative of the underlying uncertainty while simultaneously not trusting the data-driven model. We later verify this conclusion through a parameter study in the numerical experiments section.

\subsection{Adaptive Trust for data-driven models}

As the scaling factors $s_X$ and $s_U$ represent our trust in the theory-driven ensemble and data-driven models respectively, it then becomes important to determine that level of trust algorithmically. 
This section therefore outlines an algorithm for determining the bandwidth scaling factors $s_X$ and $s_U$.
Just like in section~\cref{sec:adaptive-trust} it is possible to extend the expectation maximization algorithm presented therein. 
\begin{enumerate}
    \item \textit{Expectation}: for the $i$th step of the algorithm, construct an objective function in the joint density of the prior, measurement, and bandwidth scaling factor, 
    \begin{equation}
    \begin{multlined}
        J(s_X, s_U; s_X^{(i)}, s_U^{(i)}) =\\ \mathbb{E}_{x \sim p\left(Z^+ \middle| s_X^{(i)}, s_U^{(i)}\right)} \log \left[p(x; Y = y | Z^-)p(x ; Z^- | s_X, s_U)\right],
    \end{multlined}
    \end{equation}
    where the expectation is over samples from the posterior with the  previous value of the scaling factor $s_X^{(i)}$.
    This expectation step informs us of the cost associated with maintaining our current trust. The closer this value is to a maximum the more confident we can be that the trust in both the model outputs is correct.
    \item \textit{Maximization}: for the $i$th step of the algorithm, the objective function is maximized to find the next value of the bandwidth,
    \begin{equation}
        s_X^{(i+1)} = \operatorname{arg\,max}_{s_X} J(s_X; s_X^{(i)}),
    \end{equation}
    where the the expectation in the cost is calculated through sampling.
    The maximization step adjusts our trust in both the model outputs towards the maximum of the objective function either increasing or decreasing our trust in both the outputs.
\end{enumerate}

\noindent The gradient of the cost function, $\nabla_{s_X, s_U} J(s_X; s_X^{(i)})$ we can think of as representing a change in our trust of both the theory-driven ensemble and the data-driven model. We can therefore think of the output of this value as placing us in one of the four quadrants of the guidelines outlined in the preceding section.

For a practical implementation for this work, we initially take a naive guess of the bandwidth scaling factors $s_X^{(0)}=1$ and $s_U^{(0)}=1$ as the nominal bandwidth scaling.
For sequential filtering, the initial bandwidth scaling factors are taken to be the value determined at the previous step, as we assume that our trust in the theory-guided ensemble and in the data-driven model do not change much.
Additionally, for sequential filtering, only one step of gradient ascent is performed with a small step size, as this ensures that our measures of trust are updated, but are updated slowly.

\section{Numerical Examples}
\label{eq:numerical-experiments}

We make use of two numerical examples to showcase the utility of the non-linear coupling approach.
The first numerical example is a two-dimensional static filtering problem, commonly referred to as the banana problem because of the shape of the posterior distribution.
The second test problem is a medium dimensional ($n=40$) problem commonly referred to as the Lorenz '96 equations, used for the sequential filtering example.
The Lorenz '96 equation setup is constructed such that the difference between the EnKF and true Bayesian inference is large, and at the same time such that the EnGMF requires a significant amount of particles to converge, thus requiring a new type of algorithm.

\subsection{Banana Problem}

\begin{figure}[t]
    \centering
    \includegraphics[width=0.8\linewidth]{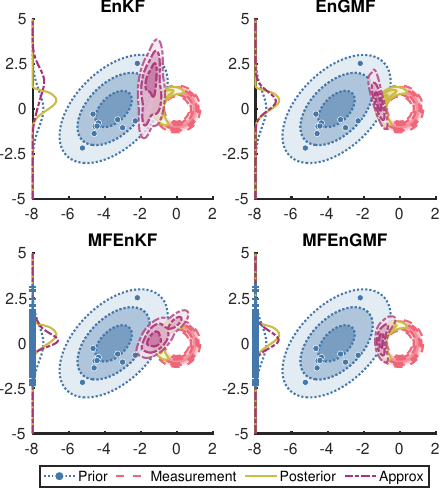}
    \caption{A visual representation of the action of four filters on the banana problem. The EnKF is on the top left, the EnGMF on the top right, the MFEnKF on the bottom left, and the MFEnKF on the bottom right. For all the panels, the blue ellipses on the left represent one, two, and three standard deviations of the prior normal distribution. The blue dots represent 10 samples therefrom. 
    The red dashed lines represent one, two, and three standard deviations of the measurement likelihood in the space of the theory-driven model. 
    The yellow solid lines contouring the banana-shape represent one, two, and three standard deviations from the true posterior distribution.
    The purple dash-dotted contours represent the posterior distributions approximated by the algorithms.
    The blue dotted line on the far left of the panels represents the prior distribution of the mock data-driven model, with the blue pluses on the lower two panels representing the data-driven samples, of which there are 100. 
    Similarly the yellow solid line  on the far left of the panels represents the true marginal of the posterior in the data-driven model space.
    The purple dash-dotted line  on the far left of the panels represents the marginal of the approximated distribution in the data-driven space.
    }
    \label{fig:banana-problem}
\end{figure}

\begin{figure}[t]
    \centering
    \begin{tikzpicture}
    \begin{axis}[clean,
        cycle list name=tol,
        xtick={10, 50, 100, 150, 200, 250},
        table/col sep=comma,
        xmin = 9,
        xmax = 251,
        ymin = -0.05,
        ymax = 5.05,
        clip = true,
        xlabel = {Theory-driven Ensemble Size ($N_X$)},
        ylabel = {Mean $f$-divergence},
        every axis plot/.append style={line width=2pt, mark size=3.5pt},
        legend style={at={(1.0,0.7)},anchor=center},
        legend cell align={left}]
    
    \addplot[mark=o,color=tolblue,dashdotted] table [x=NXs, y=kldEnMGF, col sep=comma] {bananaNKLDresults.csv};
    \addlegendentry{EnGMF ($s_X=1$)};



    \addplot[mark=x,color=tolred,dashdotted] table [x=NXs, y=kldMFEnMGF50, col sep=comma] {bananaNKLDresults.csv};
    \addlegendentry{MFEnGMF($N_U=50$, $s_X=1$, $s_U=1$)};

    \addplot[mark=o,color=tolblue] table [x=NXs, y=kldAEnMGF, col sep=comma] {bananaNKLDresults.csv};
    \addlegendentry{AEnGMF};

    \addplot[mark=x,color=tolred] table [x=NXs, y=kldAMFEnMGF50, col sep=comma] {bananaNKLDresults.csv};
    \addlegendentry{AMFEnGMF($N_U=50$)};


    \end{axis}
    \end{tikzpicture}
    \caption{Theory-driven ensemble size ($N_X$) versus an $f$-divergence for the banana problem. The two blue lines with circle markers represent the EnGMF with only the theory-driven model with the dashed line representing the EnGMF with a constant nominal scaling factor of $s_X=1$ and the solid blue line representing the adaptive trsut variant, the AEnGMF with adaptive scaling factor. The red lines with cross markers represent the MFEnGMF with both theory-driven and mock data-driven model. The dashed line represents the MFEnGMF with the theory-driven model with the nominal scaling factor $s_X=1$ and data-driven model with $N_U=50$ ensemble members and nominal scaling factor $s_U=1$. The corresponding solid red  line represents the adaptive trust version of the MFEnGM, the AEnGMF with the same models, but adaptive scaling factors.}
    \label{fig:banana-KLD1}
\end{figure}
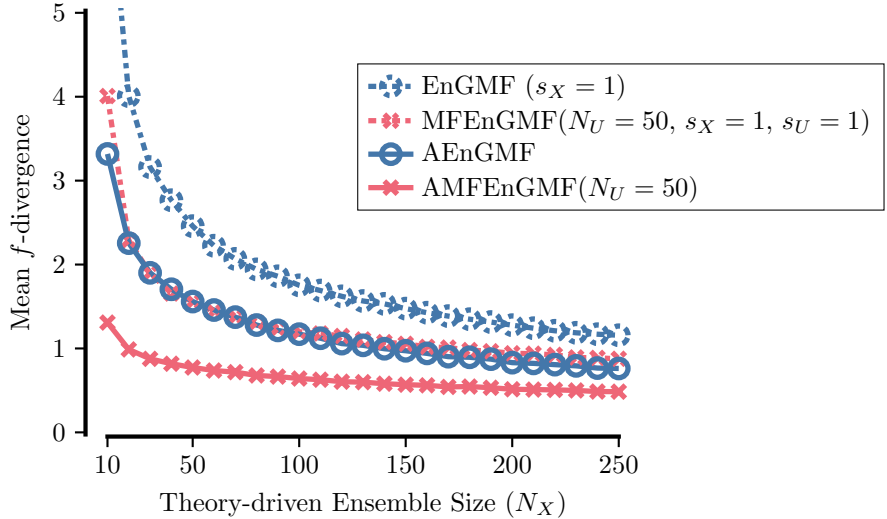

\begin{figure}[t]
    \centering
    \begin{tikzpicture}
    \begin{axis}[clean,
        cycle list name=tol,
        xtick={0, 1, 2, 3, 4, 5},
        table/col sep=comma,
        xmin = 0,
        xmax = 5.05,
        ymin = -0.05,
        ymax = 5.05,
        clip = true,
        xlabel = {Theory-driven Bandwidth Scaling ($s_X$)},
        ylabel = {Mean $f$-divergence},
        every axis plot/.append style={line width=2pt, mark size=3.5pt},
        legend style={at={(1.0,0.8)},anchor=center},
        legend cell align={left}]
    
    \addplot[mark=o,color=tolblue, dashdotted] table [x=sXas, y=kldEnMGF, col sep=comma] {bananaBandNKLDresults.csv};
    \addlegendentry{EnGMF($N_X=25$)};

    \addplot[mark=none,color=tolblue] table [x=sXas, y=kldAEnMGF, col sep=comma] {bananaBandNKLDresults.csv};
    \addlegendentry{AEnGMF($N_X=25$)};

    \addplot[mark=x,color=tolred, dashdotted] table [x=sXas, y=kldMFEnMGF1, col sep=comma] {bananaBandNKLDresults.csv};
    \addlegendentry{MFEnGMF($N_X=25$, $N_U = 50$, $s_U = 1$)};


    \addplot[mark=square,color=tolpurple, dashdotted] table [x=sXas, y=kldMFEnMGF4, col sep=comma] {bananaBandNKLDresults.csv};
    \addlegendentry{MFEnGMF($N_X=25$, $N_U = 50$, $s_U = 4$)};

    \addplot[mark=none,color=tolred] table [x=sXas, y=kldAMFEnMGF, col sep=comma] {bananaBandNKLDresults.csv};
    \addlegendentry{AMFEnGMF($N_X=25$, $N_U = 50$)};


    \end{axis}
    \end{tikzpicture}
    \caption{Theory-driven bandwidth scaling ($s_X$) versus an $f$-divergence for the banana problem. The theory-driven ensemble size is fixed to $N_X=25$. The upper solid blue horizontal line represents the AEnGMF with adaptive scaling factor, while the dashed blue line with circle markers represents the EnGMF. The dashed blue line with circle markers represents the EnGMF. The lower solid red line represents the AMFEnGMF with a data-driven ensemble size of $N_U=50$.
    The dashed red line with x markers represents the MFEnGMF with a fixed data-driven scaling factor of $s_U=1$. The dashed purple line with square markers represents the MFEnGMF with a fixed data-driven scaling factor of $s_U=4$.}
    \label{fig:banana-KLD2}
\end{figure}
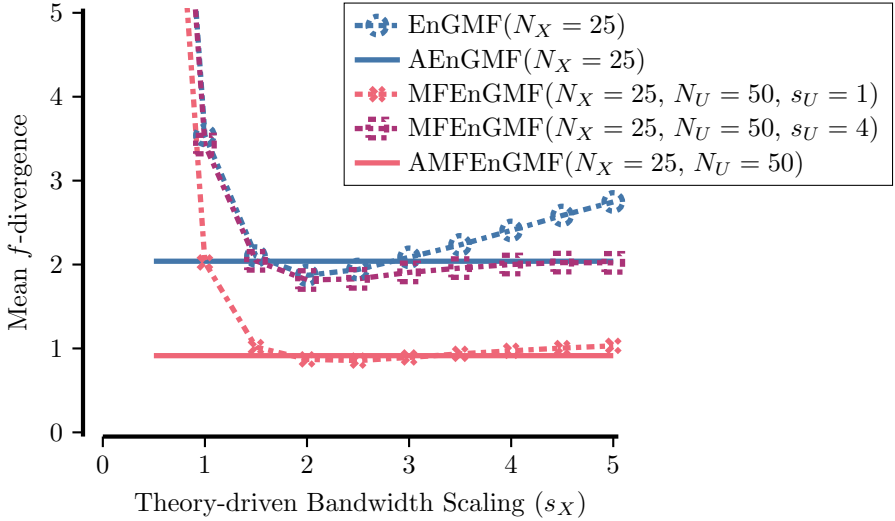

For our static filtering experiment we make use of a two-dimensional test-problem.
The prior distribution is taken to be a Gaussian with mean and covariance of,
\begin{equation}
    \mu^{X^-} = \begin{bmatrix}
        -3.5\\0
    \end{bmatrix}, \quad \Sigma^{X^-} = \begin{bmatrix}
        1 & 0.5 \\ 0.5 & 1
    \end{bmatrix},
\end{equation}
where $\mu$ is the mean and $\Sigma$ is the covariance.
The measurement function, measurement, and measurement covariance are taken to be,
\begin{equation}
    h(x) = \lVert x \rVert_2, \,\,y = 1,\,\, R = 10^{-2},
\end{equation}
meaning that there is a high degree of disagreement between the prior information and the measurement.

Instead of a real data-driven model, we take a mock data-driven model as a simple linear projection of the form,
\begin{equation}
    \theta(x) = \begin{bmatrix}
        0 & 1
    \end{bmatrix}\,\, x,
\end{equation}
which is a useful but fully interpretable surrogate.
Samples from the data-driven variate are generated by projection of samples from the theory-driven variate, meaning that the mock data-driven model is relatively trustworthy.

We first attempt to visually distinguish the action of the EnKF, MFEnKF, EnMGF, and MFEnGMF for $N_X = 10$ and $N_U = 100$ samples with nominal scaling factors of one for the latter two algorithms.
The posterior distribution of the former two algorithms is computed through kernel density estimation as the EnKF and MFEnKF are linear ensemble filters that do not make any distribution assumptions.
A visual demonstration of said action can be seen~\cref{fig:banana-problem}.
As can be seen the EnKF and MFEnKF both produce poor linear approximations to the exact posterior while the EnGMF produces a non-banana-shaped approximation. The MFEnGMF is able to start approaching the true posterior despite only 10 theory-driven ensemble members.

We next look at exclusively the non-linear filter performance for various different settings.
The metric that we look at for this work is the following $f$-divergence:
\begin{equation}
    D_f (p || q) = \int_{R^n} p(x) \left(\log p(x) - \log q(x)\right)^2 \mathrm{d}x,
\end{equation}
where $p$ is the true density of interest and $q$ is some approximation of the density of interest, and is an approximation to the KL-divergence that numerically always evaluates to a positive value.
Of note is that if $p$ and $q$ are identical, then the $f$-divergence evaluates to zero.

We look at four filters for this experiment: the EnGMF, the AEnGMF with an adaptive determination of the scaling factor $s_X$, the MFEnGMF, and the AMFEnGMF with an adaptive determination of the scaling factors $s_X$ and $s_U$.
For the adaptive filters, the AEnGMF and AMFEnGMF, one step of expectation maximization is performed with the optimization done through 25 steps of gradient ascent. The expectation in the cost function is approximated through $N_X$ samples which are kept constant throughout the optimization. This allows the gradient to be computed using forward finite difference with a step of $10^{-6}$.
The initial guesses for both $s_X$ and $s_U$ are fixed to one, and the step size is fixed to be $1/32$ for both AEnGMF and the AMFEnGMF, which was hand-tuned to maximize numerical stability.
All experiments are run for $10,008$ Monte Carlo iterations (as to divide by the 36 physical cores of the computer on which the experiments were performed) and $f$-divergence averaged over said iterations.

The first experiment that we run for these four filters is that of varying the theory-driven ensemble size $N_X$. The scaling factor is only varied for the EnGMF and the MFEnMGF, while for the adaptive filters nothing is varied.
For the two multifidelity algorithms the data-driven ensemble size is fixed to $N_U=50$.
The results for this experiment can be seen in~\cref{fig:banana-KLD1}. As can be seen, the EnGMF with fixed bandwidth scaling factor $s_X = 1$ is by far the worst performing algorithm. Next, the EnGMF with adaptive trust, the AEnGMF and the MFEnGMF with fixed bandwidth scaling factors are roughly equal in performance. This indicates that including a mock data-driven model in the inference roughly confers the same reduction in $f$-divergence as tuning our trust in the theory-driven ensemble. By far the best algorithm is that of the adaptive trust MFEnGMF, the AMFEnGMF. 
For $N_X=10$ theory-driven ensemble members it matches the $f$-divergence that the AEnGMF has for $N_X=100$ theory-guided ensemble members.  
These results signal that the AMFEnGMF accurately merges the theory-driven and data-driven models in a way that is greater than the sum of their parts.

The second experiment that we run for these filters is that of varying the theory-driven scaling factor $s_X$. The scaling factor is only varied for the EnGMF and the MFEnMGF, while for the adaptive filters nothing is varied.
Again, for the two multifidelity algorithms the data-driven ensemble size is fixed to $N_U=50$.
The MFEnGMF is additionally run with two different data-driven scaling factors of $s_U = 1$ and $s_U = 4$. 
The theory-driven ensemble size is fixed to $N_X = 25$.
The results for this experiment can be seen in~\cref{fig:banana-KLD2}. As can be seen, the EnGMF performs worse than either the MFEnGMF and the AMFEnGMF. The adaptive trust EnGMF performs almost as good as the EnGMF with the lowest-error theory-driven bandwidth of $s_X = 2$, meaning that the adaptive algorithm performs nearly close to optimal without having direct access to the truth.
For the MFEnGMF with $s_U=4$, the case where the data-driven model is not trusted that much, the results clearly indicate that the $f$-divergence is close to that of the EnGMF, meaning that when the data-driven model is not trusted, the algorithm reverts to the EnGMF, as expected.

\subsection{Lorenz '96}

The $40$-variable Lorenz '96 problem~\cite{lorenz1996predictability,van2018dynamics} is defined as a system of ordinary differential equations
\begin{equation}
    x_k' = -x_{k-1}(x_{k-2} - x_{k+1}) - x_k + F,\quad k = 1,\dots, 40,
\end{equation}
with cyclic boundary conditions, $x_0 = x_{40}$, $x_{-1} = x_{39}$, and $x_{41} = x_1$, as is canonical.
The forcing factor is fixed for $F = 8$ which makes the dynamical system chaotic with Kaplan-Yorke dimension of $27.1$~\cite{popov2019bayesian}.
The initial condition is taken to lie on the attractor, as is typical, but whose value does not change the behavior of the system.
Measurements are performed every $0.2$ time units, which does not correspond to any physical time, as the system does not correspond to any discretization of a physical dynamical system. 
The time integration is performed with a fourth order Runge-Kutta method with a fixed step size of $0.05$.

\subsubsection{Na\"ive AI-driven reduced order model}

In this section we construct a neural-network-based (artificial intelligence-based) data-driven model of the Lorenz '96 equations through an autoencoder-based reduced order model.
The reduced order model is deliberately constructed in a very naive fashion in order for it to be deliberately untrustworthy in order to attempt to showcase a worst-case scenario.

We make use of a right-invertible autoencoder~\cite{popov2022multifidelity} framework to create the reduced order model for the Lorenz '96 system.
The dimensionality reduction is performed as follows,
\begin{equation}
    \theta(\phi(x)) \approx x,\quad  \phi(\theta(\phi(x))) \approx \phi(x),
\end{equation}
where $\phi$ is the encoder and $\theta$ is the decoder, and the right-invertible constraint is weakly enforced with a factor of $\lambda = 10$. The encoder and the decoder were both taken to be a one hidden layer fully-connected network with hidden dimension of $100$ for both and a $\tanh$ non-linearity (erroneously known as the activation function).
Three distinct dimensions of the reduced system are taken, $r\in\{14,\,28\,35\}$, with $14$ representing the number of positive Lyapunov modes, $28$ representing the integer ceiling of the Kaplan-Yorke dimension and $35$ representing the dimension of the data~\cite{popov2022multifidelity}. The number of data points collected from the original system is taken to be only $2000$, again to deliberately create untrustworthy models. 

For neural network optimization, the Adam~\cite{kingma2014adam} algorithm is used with a batch-size the same as the data size, thus making the algorithm deterministic.
The hyperparamters of the algorithm are set to $\beta_1=0.9$ which is the parameter controlling the momentum term, $\beta_2=0.95$ which is the parameter controlling the scaling of each term (note that as we are performing deterministic optimization $\beta_2$ is significantly smaller than is used for standard applications with smaller batch-sizes), and $\epsilon=10^{-8}$ which is a numerical stability term.
A triangular step-size scheduler with a minimum step-size of $10^{-4}$ and a maximum step-size of $10^{-2}$  was used.

As constructing an optimal reduced order model is not the focus of this work, the data-driven forward model is defined in terms of the autoencoder and the theory-driven forward model, na\"ively as follows:
\begin{equation}
    \mathcal{M}_U(u) = \phi(\mathcal{M}_X(\theta(u))) + \eta_U,
\end{equation}
meaning that $u$ is projected up from the space of the data-driven model to the space of the theory-driven model, propagated forwards, and then projected back down.
The error term $\eta_U$ is taken to be normally distributed with empirically computed mean and covariance over the data points.

The laissez-faire way in which the data-driven model is constructed is on purpose. We wish to build a model that does not have any guard rails and that we can guarantee will be inaccurate and fail. Despite all this, we show that the model can be useful.

\subsubsection{Sequential filtering experiments}

\begin{table}[t]
    \centering
\begin{tabularx}{0.75\linewidth}{Xcccccc}\toprule
                                  & $N_X = $   & $5$   & $10$   & $25$  & $50$   & $100$  \\\toprule
EnGMF                             & $s_X$      & $1.0$ & $1.0$  & $1.5$ & $1.5$  & $1.5$  \\\hline
\multirow{2}{*}{MFEnGMF ($r=14$)} & $s_X$      & $3.0$ & $2.0$  & $1.5$ & $1.5$  & $1.5$  \\
                                  & $s_U$      & $2.5$ & $1.5$  & $1.5$ & $1.5$  & $1.5$  \\\hline
\multirow{2}{*}{MFEnGMF ($r=28$)} & $s_X$      & $3.0$ & $2.5$  & $2.0$ & $1.5$  & $1.5$  \\
                                  & $s_U$      & $1.0$ & $2.0$  & $2.0$ & $2.5$  & $2.0$  \\\hline
\multirow{2}{*}{MFEnGMF ($r=35$)} & $s_X$      & $3.0$ & $3.0$  & $2.0$ & $1.5$  & $1.5$  \\
                                  & $s_U$      & $1.0$ & $2.0$  & $2.0$ & $1.5$  & $1.5$  \\\toprule
EnKF                              & $\alpha_X$ & $1.1$ & $1.1$  & $1.1$ & $1.1$  & $1.1$  \\\hline
\multirow{2}{*}{MFEnKF ($r=14$)}  & $\alpha_X$ & $1.1$ & $1.1$  & $1.1$ & $1.1$  & $1.1$  \\
                                  & $\alpha_U$ & $1.1$ & $1.1$  & $1.1$ & $1.1$  & $1.1$  \\\hline
\multirow{2}{*}{MFEnKF ($r=28$)}  & $\alpha_X$ & $1.1$ & $1.1$  & $1.1$ & $1.1$  & $1.1$  \\
                                  & $\alpha_U$ & $1.1$ & $1.1$  & $1.1$ & $1.1$  & $1.1$  \\\hline
\multirow{2}{*}{MFEnKF ($r=35$)}  & $\alpha_X$ & $1.1$ & $1.1$  & $1.1$ & $1.1$  & $1.1$  \\
                                  & $\alpha_U$ & $1.1$ & $1.1$  & $1.1$ & $1.1$  & $1.1$  \\\hline
\end{tabularx}
    \caption{Results of parameter study for the Lorenz '96 equations. The top part of the table provides the optimal parameter values in terms of scaling factor $s_X$ for the theory-driven Gaussian mixture kernel density estimate, and, where applicable, the scaling factor $s_U$ for the data-driven Gaussian mixture kernel density estimate.}
    \label{tab:lorenz96-parameter-study}
\end{table}

\begin{figure}[t]
    \centering
    \begin{tikzpicture}
    \begin{axis}[clean,
        cycle list name=tol,
        xmode=log,
        log ticks with fixed point,
        xtick={5, 10, 25, 50, 100},
        table/col sep=comma,
        xmin = 4,
        xmax = 120,
        ymin = -0.05,
        ymax = 5.05,
        clip = true,
        xlabel = {Theory-driven Ensemble Size ($N_X$)},
        ylabel = {Mean Spatio-temporal RMSE},
        every axis plot/.append style={line width=2pt, mark size=3.5pt},
        legend style={at={(1.30,0.7)},anchor=center},
        legend cell align={left}]
    
    \addplot[mark=o,color=tolblue, dashdotted] table [x=NXs, y=rmseEnGMF, col sep=comma] {fullrmse.csv};
    \addlegendentry{EnGMF};

    \addplot[mark=o,color=tolblue] table [x=NXs, y=rmseAEnGMF, col sep=comma] {adaptivermse.csv};
    \addlegendentry{AEnGMF};

    \addplot[mark=square,color=tolpurple] table [x=NXs, y=rmseEnKF, col sep=comma] {fullrmse.csv};
    \addlegendentry{EnKF};

    \addplot[mark=x,color=tolred, dashdotted] table [x=NXs, y=rmseMFEnGMF14, col sep=comma] {fullrmse.csv};
    \addlegendentry{MFEnGMF($r=14$)};

    \addplot[mark=x,color=tolred] table [x=NXs, y=rmseAMFEnGMF14, col sep=comma] {adaptivermse.csv};
    \addlegendentry{AMFEnGMF($r=14$)};

    \addplot[mark=+,color=tolyellow] table [x=NXs, y=rmseMFEnKF14, col sep=comma] {fullrmse.csv};
    \addlegendentry{MFEnKF($r=14$)};

    \addplot[mark=none,color=black, dashed, domain=5:100]  {0.5413};
    \addlegendentry{Bayesian Inference};

    \addplot[mark=none,color=black, domain=5:100]  {3.6269};
    \addlegendentry{No Filter};
    
    \end{axis}
    \end{tikzpicture}
    \caption{Theory-driven ensemble size ($N_X$) versus mean spatio-temporal RMSE for the Lorenz '96 equations. The upper solid black horizontal line represents the error if all observations are ignored, thus no filtering is performed. The lower dashed horizontal black line represents an approximation to true Bayesian inference using $N=10000$ ensemble members of an EnGMF, and is close to the theoretically lowest error possible. The blue lines with circle markers represent the EnGMF and AEnGMF, with dashed and solid lines respectfully. The purple line with square markers represents the EnKF, which is the state-of-the-art linear filter. The red lines with x markers represent the MFEnGMF and AMFEnGMF, with dashed and solid lines respectfully with the reduced order dimension taken to be $r=14$ and the reduced order ensemble size taken to be $N_U=100$. The yellow line with plus markers represents the MFEnKF with the same settings as the other multi-fidelity filters.}
    \label{fig:lorenz96-experiment-r14}
\end{figure}
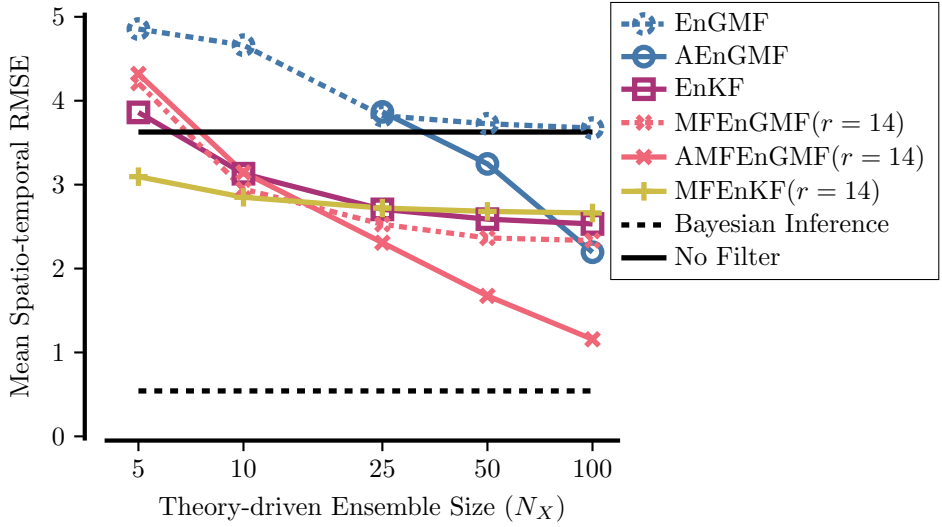

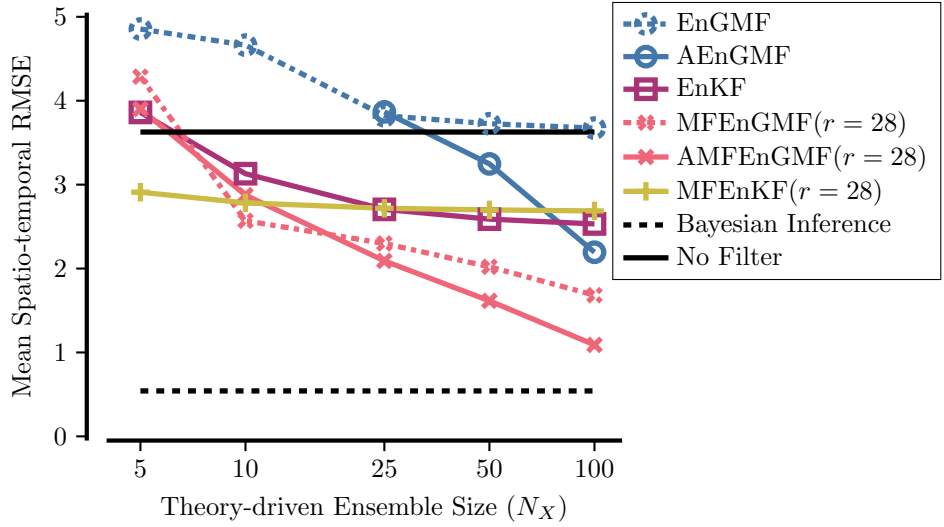
\begin{figure}[t]
    \centering
    \begin{tikzpicture}
    \begin{axis}[clean,
        cycle list name=tol,
        xmode=log,
        log ticks with fixed point,
        xtick={5, 10, 25, 50, 100},
        table/col sep=comma,
        xmin = 4,
        xmax = 120,
        ymin = -0.05,
        ymax = 5.05,
        clip = true,
        xlabel = {Theory-driven Ensemble Size ($N_X$)},
        ylabel = {Mean Spatio-temporal RMSE},
        every axis plot/.append style={line width=2pt, mark size=3.5pt},
        legend style={at={(1.30,0.7)},anchor=center},
        legend cell align={left}]
    
    \addplot[mark=o,color=tolblue, dashdotted] table [x=NXs, y=rmseEnGMF, col sep=comma] {fullrmse.csv};
    \addlegendentry{EnGMF};

    \addplot[mark=o,color=tolblue] table [x=NXs, y=rmseAEnGMF, col sep=comma] {adaptivermse.csv};
    \addlegendentry{AEnGMF};

    \addplot[mark=square,color=tolpurple] table [x=NXs, y=rmseEnKF, col sep=comma] {fullrmse.csv};
    \addlegendentry{EnKF};

    \addplot[mark=x,color=tolred, dashdotted] table [x=NXs, y=rmseMFEnGMF28, col sep=comma] {fullrmse.csv};
    \addlegendentry{MFEnGMF($r=28$)};

    \addplot[mark=x,color=tolred] table [x=NXs, y=rmseAMFEnGMF28, col sep=comma] {adaptivermse.csv};
    \addlegendentry{AMFEnGMF($r=28$)};

    \addplot[mark=+,color=tolyellow] table [x=NXs, y=rmseMFEnKF28, col sep=comma] {fullrmse.csv};
    \addlegendentry{MFEnKF($r=28$)};

    \addplot[mark=none,color=black, dashed, domain=5:100]  {0.5413};
    \addlegendentry{Bayesian Inference};

    \addplot[mark=none,color=black, domain=5:100]  {3.6269};
    \addlegendentry{No Filter};
    
    \end{axis}
    \end{tikzpicture}
    \caption{Similar to~\cref{fig:lorenz96-experiment-r14}, but with the reduced order dimension taken to be $r=28$.}
    \label{fig:lorenz96-experiment-r28}
\end{figure}

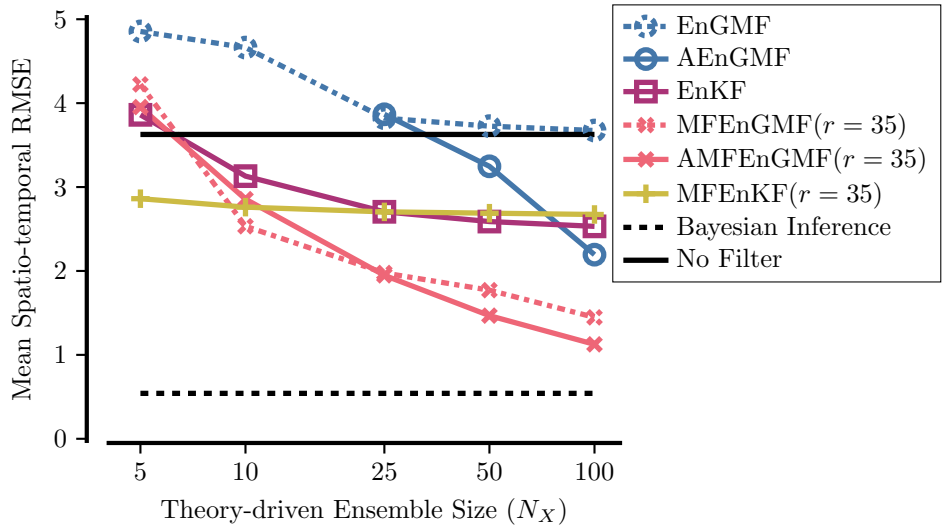
\begin{figure}[t]
    \centering
    \begin{tikzpicture}
    \begin{axis}[clean,
        cycle list name=tol,
        xmode=log,
        log ticks with fixed point,
        xtick={5, 10, 25, 50, 100},
        table/col sep=comma,
        xmin = 4,
        xmax = 120,
        ymin = -0.05,
        ymax = 5.05,
        clip = true,
        xlabel = {Theory-driven Ensemble Size ($N_X$)},
        ylabel = {Mean Spatio-temporal RMSE},
        every axis plot/.append style={line width=2pt, mark size=3.5pt},
        legend style={at={(1.30,0.7)},anchor=center},
        legend cell align={left}]
    
    \addplot[mark=o,color=tolblue, dashdotted] table [x=NXs, y=rmseEnGMF, col sep=comma] {fullrmse.csv};
    \addlegendentry{EnGMF};

    \addplot[mark=o,color=tolblue] table [x=NXs, y=rmseAEnGMF, col sep=comma] {adaptivermse.csv};
    \addlegendentry{AEnGMF};

    \addplot[mark=square,color=tolpurple] table [x=NXs, y=rmseEnKF, col sep=comma] {fullrmse.csv};
    \addlegendentry{EnKF};

    \addplot[mark=x,color=tolred, dashdotted] table [x=NXs, y=rmseMFEnGMF35, col sep=comma] {fullrmse.csv};
    \addlegendentry{MFEnGMF($r=35$)};

    \addplot[mark=x,color=tolred] table [x=NXs, y=rmseAMFEnGMF35, col sep=comma] {adaptivermse.csv};
    \addlegendentry{AMFEnGMF($r=35$)};

    \addplot[mark=+,color=tolyellow] table [x=NXs, y=rmseMFEnKF35, col sep=comma] {fullrmse.csv};
    \addlegendentry{MFEnKF($r=35$)};

    \addplot[mark=none,color=black, dashed, domain=5:100]  {0.5413};
    \addlegendentry{Bayesian Inference};

    \addplot[mark=none,color=black, domain=5:100]  {3.6269};
    \addlegendentry{No Filter};
    
    \end{axis}
    \end{tikzpicture}
    \caption{Similar to~\cref{fig:lorenz96-experiment-r14}, but with the reduced order dimension taken to be $r=35$.}
    \label{fig:lorenz96-experiment-r35}
\end{figure}

In this section we run sequential filtering example for the Lorenz '96 equations.
We run over 1100 observations of the Lorenz '96 equations with observations occurring every $0.2$ time units.
The equations are evolved using a fourth order Runge-Kutta method with a fixed time step of $0.05$.
The observations are taken to come from the non-linear observation operator,
\begin{equation}
    \left[h(x)\right]_i = \sqrt{x_{2i-1}^2 + x_{2i}^2},\quad i = 1, \dots, 20,
\end{equation}
with unbiased (mean zero) Gaussian observation error with covariance $R = 0.25 I_20$.

The first 100 steps of the algorithms are discarded for spinup (burn-in) and the mean spatio-temporal root-mean-squared-error between the posterior mean estimate $\bar{x}$, and truth $x^t$ (over $T=1000$ time units of interest, with dimension $n=40$),
\begin{equation}\label{eq:RMSE}
    \operatorname{RMSE}(\bar{x}, x^t) = \sqrt{\frac{1}{nT}\sum_{i=1}^T\sum_{l=1}^n \left(\bar{x}_{i,l} - x^t_{i,l}\right)^2},
\end{equation}
is averaged over 20 Monte Carlo iterations where the initial ensembles, and observations are all varied.

We run the sequential filtering experiments on the following filters:
the EnGMF, the EnKF, the MFEnKF, the MFEnGMF and the adaptive filters, the AEnGMF and AMFEnGMF.
For the multifidelity filters (the MFEnKF, MFEnMGF, AMFEnMGF) a fixed data-driven ensemble size of $N_U=100$ was used, and the reduced order dimensions are taken to be $r\in\{14,28,35\}$.
All filters are run for a variety of theory-driven ensemble sizes of $N_X \in \{5, 10, 25, 50, 100\}$.
The particular scenario of note is when the theory-driven ensemble size is less than the dimension of the system, which is $n=40$, meaning that of particular interest are the cases $N_X = 5$, $N_X=10$, and $N_X = 25$.

In order to examine the ideal cases of the non-adaptive algorithms, for the EnGMF, MFEnGMF, EnKF, and MFEnKF we assume that we have an oracle that determines the best possible set of static parameters. In effect for these algorithms we perform a parameter sweep over the set of all possible parameters and choose
The goal of the parameter sweep in the case of the EnGMF and MFEnGMF algorithms is to show the interaction between the different scaling factors and to show how the scaling factors can be viewed as proxies for our measures of trust in the ensembles and models~(\cref{sec:scaling-factor-discussion,sec:trust}).
As a parameter sweep is unrealistic in reality, these algorithms are also provided to show a baseline which the adaptive algorithms strive to match or even beat.
For the EnGMF and MFEnGMF, both scaling factors are swept over the sets $s_X,\, s_U\in\{0.5, 1, 1.5, 2, 2.5, 3\}$, and for the EnKF and MFEnKF the inflation factors are swept over the sets $\alpha_X,\, \alpha_U\in\{1, 1.05, 1.1\}$.

For the adaptive algorithms, the AEnGMF, and the AMFEnGMF, the initial scaling factors are set to $s_X^{(0)} = 1$ and $s_U^{(0)} = 1$, meaning that our initial trust in the theory-driven ensemble and data-driven model is nominal. 
At each step, the initial estimate of the scaling factors is taken from the previous step.
Subsequently, one step of expectation maximization is performed with one step of gradient ascent with a step size of $1/128$ for numerical stability. In essence we assume that our trust in the theory-guided ensemble and data-driven model do not change much from the previous step, but are still adaptively updated to better match the current observations.

For all algorithms, for computing the covariance, B-localization~\cite{asch2016data} was performed,
\begin{equation}
    \Sigma^X \leftarrow \rho \circ \Sigma^X,
\end{equation}
where $\circ$ is an element-wise product and the matrix $\rho$ is defined  through a Gaussian decorrelation function with radius $r=4$,
\begin{equation}
\begin{gathered}
    \rho_{i,j} = \operatorname{exp}\left(-\frac{1}{2} \frac{d(i, j)^2}{r^2}\right),\quad i,j = 1, \dots 40,\\
    d(i,j) = \operatorname{min}\{ \lvert i - j \rvert, \lvert 40 + i - j \rvert, \lvert 40 + j - i \rvert\}
\end{gathered}
\end{equation}
where $d$ is the index-wise distance function defined on the Lorenz '96 equations.

An approximation to the theoretically minimal error through (a good approximation to) exact Bayesian inference was computed by making use of an EnGMF with $N=10^5$ particles.
In order to see if filtering inhibits reliable estimation, a `no filter' scenario was also computed, where only forecasts were performed with no analysis.

We first look at the results of the parameter sweep in table~\cref{tab:lorenz96-parameter-study}. It can be seen that all EnKF-derived algorithms prefer a larger inflation factor for the given problem, though as the algorithms are not convergent it is the authors opinion that even larger values would not have a significant impact on the error.
For the EnGMF, the scaling factor $s_X = 1.5$ is preferred to all others, though as will be seen later, this does not have an impact on the convergence of the algorithm.
As there is no other model to compare to, this factor as a proxy for our trust seems to have limited use.
For the MFEnGMF, there is an interesting spectrum of scaling factors. In the case of $N_X=5$, the scaling factor $s_X$ for the theory-guided ensemble is always high with the maximum value of $s_X = 3.0$ having the lowest error. This means that, as the ensemble size is small, there is little trust in its reliability as opposed to the data-driven model.
Furthermore, looking at the case $r=14$, there also seems to be little trust in the data-driven model as well when compared to the cases $r=28$ and $r=35$ for which there is more trust in the data-driven models.
It can be seen that as the theory-guided ensemble size $N_X$ increases, the trust theory-guided ensemble increases through a decrease in the theory-guided scaling factor $s_X$, while the trust in the data-driven model seems to stay constant past $N_X=10$ as evidenced by very little change in the data-driven scaling factor $s_U$. In the author's view this validates the ideas discussed in~\cref{sec:trust}.

We then look at the sequential filtering results for $r=14$ in~\cref{fig:lorenz96-experiment-r14}, for $r=28$ in~\cref{fig:lorenz96-experiment-r28} and for $r=35$ in~\cref{fig:lorenz96-experiment-r35}.
As can be seen, the EnKF performs better than the `no filter' scenario in all cases except for $N_X=5$ for which case the best filter is always the MFEnKF. Both the EnKF and the MFEnKF reach a point after which an increase in the ensemble size does not provide a decrease in the error as the filters are not convergent. 
The EnGMF, even with an oracle determining the optimal scaling factor never achieves error below that of the `no filter' scenario for the small theory-driven ensemble sizes. The EnGMF does provide for a significant decrease in erorr for $N_X=10^5$, which is the `Bayesian inference' approximation, though in a real-world scenario that cost would be too high.
The AEnGMF, where the scaling factor is adaptively determined, the EnGMF always requires more theory-driven ensemble members than the dimension of the system $n=40$ in order to perform better than the `no filter' scenario.
The MFEnGMF, with an oracle determining our trust in the theory-driven ensemble and in the data-driven model, performs as good, or superior to the MFEnKF for as few theory-driven ensemble members as $N_X=10$, with larger models perfoming better.
Finally we have the AMFEnGMF, which adaptively determines our trust in the theory-driven ensemble and in the data-driven model through their respective scaling factors, without the need for an oracle. For all reduced order models the AMFEnGMF performs at least as well as the EnKF for $N_X=10$, and, most importantly, is always the best filter for the undersampled scenario $N_X=25$, even for the worst data-driven reduced order model with dimension $r=14$. This trend continues with the AMFEnGMF fast approaching the `Bayesian inference' scenario at $N_X=100$.

The results above conclusively show that it is possible to trust an artificial intelligence or data-driven model for data assimilation purposes, provided that our level of trust in that model is regulated, and that the model is not fully trusted implicitly.
Moreover,  the above results validate the idea that it is possible to adaptively determine our relative trust between the results of a theory-guided model and a data-driven model in a multifidelity ensemble Gaussian mixture filtering framework by adaptively determining the bandwidth scaling factors of the underlying kernel density estimations. 

\section{Conclusions and future outlook}
\label{sec:conclusions}

This work provides a new method for learning to trust AI-driven and data-driven models for data assimilation.
We have shown that our relative trust in both the theory-driven and data-driven models can be represented by 
by leveraging a theory-driven principal model in a multifidelity ensemble Gaussian mixture framework.
We demonstrated that it is possible to adaptively determine our trust in both the theory-driven and data-driven models.
Most importantly, we have shown that this strategy can potentially create a convergent particle filter for a high-dimensional 40-variable system, the Lorenz '96 equations.
Future work will combine the multifidelity ensemble Kalman filter and the multifidelity ensemble Gaussian mixture filter for very high dimensional data assimilation problems.
Other future work would extend this idea to the ensemble Epanechnikov mixture filter~\cite{popov2024Epanechnikov}.

\section*{Acknowledgments}
The author would like to thank the University of Hawai'i for startup funds that helped fund this work.

\bibliographystyle{siamplain}
\bibliography{biblio}

@article{popov2021multifidelity,
  title={A multifidelity ensemble {Kalman} filter with reduced order control variates},
  author={Popov, Andrey A and Mou, Changhong and Sandu, Adrian and Iliescu, Traian},
  journal={SIAM Journal on Scientific Computing},
  volume={43},
  number={2},
  pages={A1134--A1162},
  year={2021},
  publisher={SIAM}
}

@incollection{popov2022multifidelityDA,
  title={Multifidelity data assimilation for physical systems},
  author={Popov, Andrey A and Sandu, Adrian},
  booktitle={Data Assimilation for Atmospheric, Oceanic and Hydrologic Applications (Vol. IV)},
  pages={43--67},
  year={2022},
  publisher={Springer}
}

@article{popov2022modelforest,
  title={The Model Forest Ensemble {Kalman} Filter},
  author={Popov, Andrey A and Sandu, Adrian},
  journal={arXiv preprint arXiv:2210.11971},
  year={2022}
}

@article{durant2025kernel,
  title={Kernel-Based Ensemble Gaussian Mixture Probability Hypothesis Density Filter},
  author={Durant, Dalton and Zanetti, Renato},
  journal={arXiv preprint arXiv:2505.00131},
  year={2025}
}

@article{popov2020explicit,
  title={An explicit probabilistic derivation of inflation in a scalar ensemble {Kalman} filter for finite step, finite ensemble convergence},
  author={Popov, Andrey A and Sandu, Adrian},
  journal={arXiv preprint arXiv:2003.13162},
  year={2020}
}

@article{evensen1994sequential,
  title={Sequential data assimilation with a nonlinear quasi-geostrophic model using Monte Carlo methods to forecast error statistics},
  author={Evensen, Geir},
  journal={Journal of Geophysical Research: Oceans},
  volume={99},
  number={C5},
  pages={10143--10162},
  year={1994},
  publisher={Wiley Online Library}
}

@book{kalnay2003atmospheric,
  title={Atmospheric modeling, data assimilation and predictability},
  author={Kalnay, Eugenia},
  year={2003},
  publisher={Cambridge university press}
}

@article{kalman1960new,
    author = {Kalman, R. E.},
    title = {A New Approach to Linear Filtering and Prediction Problems},
    journal = {Journal of Basic Engineering},
    volume = {82},
    number = {1},
    pages = {35-45},
    year = {1960},
    month = {03},
    issn = {0021-9223},
    doi = {10.1115/1.3662552},
    url = {https://doi.org/10.1115/1.3662552},
    eprint = {https://asmedigitalcollection.asme.org/fluidsengineering/article-pdf/82/1/35/5518977/35\_1.pdf},
}

@article{kingma2014adam,
  title={Adam: A method for stochastic optimization},
  author={Kingma, Diederik P},
  journal={arXiv preprint arXiv:1412.6980},
  year={2014}
}

@book{jaynes2003probability,
  title={Probability theory: The logic of science},
  author={Jaynes, Edwin T},
  year={2003},
  publisher={Cambridge university press}
}

@book{reich2015probabilistic,
  title={Probabilistic forecasting and Bayesian data assimilation},
  author={Reich, Sebastian and Cotter, Colin},
  year={2015},
  publisher={Cambridge University Press}
}

@article{anderson1999monte,
  title={A {Monte} {Carlo} implementation of the nonlinear filtering problem to produce ensemble assimilations and forecasts},
  author={Anderson, Jeffrey L and Anderson, Stephen L},
  journal={Monthly weather review},
  volume={127},
  number={12},
  pages={2741--2758},
  year={1999}
}

@article{liu2016efficient,
  title={Efficient kernel-based ensemble {Gaussian} mixture filtering},
  author={Liu, Bo and Ait-El-Fquih, Boujemaa and Hoteit, Ibrahim},
  journal={Monthly Weather Review},
  volume={144},
  number={2},
  pages={781--800},
  year={2016}
}

@article{burgers1998analysis,
  title={Analysis scheme in the ensemble {Kalman} filter},
  author={Burgers, Gerrit and Jan van Leeuwen, Peter and Evensen, Geir},
  journal={Monthly weather review},
  volume={126},
  number={6},
  pages={1719--1724},
  year={1998}
}

@article{yun2022kernel,
  title={Kernel-based ensemble {Gaussian} mixture filtering for orbit determination with sparse data},
  author={Yun, Sehyun and Zanetti, Renato and Jones, Brandon A},
  journal={Advances in Space Research},
  volume={69},
  number={12},
  pages={4179--4197},
  year={2022},
  publisher={Elsevier}
}

@article{popov2024adaptive,
  title={An adaptive covariance parameterization technique for the ensemble Gaussian mixture filter},
  author={Popov, Andrey A and Zanetti, Renato},
  journal={SIAM Journal on Scientific Computing},
  volume={46},
  number={3},
  pages={A1949--A1971},
  year={2024},
  publisher={SIAM}
}

@book{silverman2018density,
  title={Density estimation for statistics and data analysis},
  author={Silverman, Bernard W},
  year={2018},
  publisher={Routledge}
}

@article{popov2024Epanechnikov,
  title={The Ensemble Epanechnikov Mixture Filter},
  author={Popov, Andrey A and Zanetti, Renato},
  journal={arXiv preprint arXiv:2408.11164},
  year={2024}
}

@book{asch2016data,
  title={Data assimilation: methods, algorithms, and applications},
  author={Asch, Mark and Bocquet, Marc and Nodet, Ma{\"e}lle},
  year={2016},
  publisher={SIAM}
}

@article{popov2019bayesian,
  title={A Bayesian approach to multivariate adaptive localization in ensemble-based data assimilation with time-dependent extensions},
  author={Popov, Andrey A and Sandu, Adrian},
  journal={Nonlinear Processes in Geophysics},
  volume={26},
  number={2},
  pages={109--122},
  year={2019},
  publisher={Copernicus Publications G{\"o}ttingen, Germany}
}

@inproceedings{lorenz1996predictability,
  title={Predictability: A problem partly solved},
  author={Lorenz, Edward N},
  booktitle={Proc. Seminar on predictability},
  volume={1},
  pages={1--18},
  year={1996},
  organization={Reading}
}

@phdthesis{van2018dynamics,
title = "Dynamics of the Lorenz-96 model: Bifurcations, symmetries and waves",
author = "{van Kekem}, {Dirk Leendert}",
year = "2018",
language = "English",
isbn = "978-94-034-0979-5",
publisher = "Rijksuniversiteit Groningen",
school = "University of Groningen",

}

@article{popov2022multifidelity,
  title={Multifidelity ensemble {Kalman} filtering using surrogate models defined by theory-guided autoencoders},
  author={Popov, Andrey A and Sandu, Adrian},
  journal={Data-driven modeling and optimization in fluid dynamics: From physics-based to machine learning approaches},
  volume={16648714},
  pages={41},
  year={2023},
  publisher={Frontiers Media SA}
}

@article{hicks2024chatgpt,
  title={ChatGPT is bullshit},
  author={Hicks, Michael Townsen and Humphries, James and Slater, Joe},
  journal={Ethics and Information Technology},
  volume={26},
  number={2},
  pages={1--10},
  year={2024},
  publisher={Springer}
}

@article{cheetham2024artificial,
  title={Artificial intelligence driving materials discovery? perspective on the article: Scaling deep learning for materials discovery},
  author={Cheetham, Anthony K and Seshadri, Ram},
  journal={Chemistry of Materials},
  volume={36},
  number={8},
  pages={3490--3495},
  year={2024},
  publisher={ACS Publications}
}

@article{kapoor_leakage_2023,
  title = {Leakage and the reproducibility crisis in machine-learning-based science},
  issn = {2666-3899},
  url = {https://www.cell.com/patterns/abstract/S2666-3899(23)00159-9},
  doi = {10.1016/j.patter.2023.100804},
  language = {English},
  journal = {Patterns},
  author = {Kapoor, Sayash and Narayanan, Arvind},
  month = aug,
  year = {2023},
  note = {Publisher: Elsevier},
}

@article{mcgreivy2024weak,
  title={Weak baselines and reporting biases lead to overoptimism in machine learning for fluid-related partial differential equations},
  author={McGreivy, Nick and Hakim, Ammar},
  journal={Nature machine intelligence},
  volume={6},
  number={10},
  pages={1256--1269},
  year={2024},
  publisher={Nature Publishing Group UK London}
}

@inproceedings{michaelson2023ensemble,
  title={Ensemble {Kalman} filter with Bayesian recursive update},
  author={Michaelson, Kristen and Popov, Andrey A and Zanetti, Renato},
  booktitle={2023 26th International Conference on Information Fusion (FUSION)},
  pages={1--6},
  year={2023},
  organization={IEEE}
}

@article{bocquet2020bayesian,
  title={Bayesian inference of chaotic dynamics by merging data assimilation, machine learning and expectation-maximization},
  author={Bocquet, Marc and Brajard, Julien and Carrassi, Alberto and Bertino, Laurent},
  journal={arXiv preprint arXiv:2001.06270},
  year={2020}
}

@incollection{neal1998view,
  title={A view of the EM algorithm that justifies incremental, sparse, and other variants},
  author={Neal, Radford M and Hinton, Geoffrey E},
  booktitle={Learning in graphical models},
  pages={355--368},
  year={1998},
  publisher={Springer}
}

@book{bishop2006pattern,
  title={Pattern recognition and machine learning},
  author={Bishop, Christopher M and Nasrabadi, Nasser M},
  volume={4},
  year={2006},
  publisher={Springer}
}

@misc{popov2026divideconquerstrategymultinomial,
      title={A divide and conquer strategy for multinomial particle filter resampling}, 
      author={Andrey A. Popov},
      year={2026},
      eprint={2604.01356},
      archivePrefix={arXiv},
      primaryClass={cs.DS},
      url={https://arxiv.org/abs/2604.01356}, 
}

@book{jsang2018subjective,
  title={Subjective Logic: A formalism for reasoning under uncertainty},
  author={Jsang, Audun},
  year={2018},
  publisher={Springer Publishing Company, Incorporated}
}

@book{shafer2020mathematical,
  title={A mathematical theory of evidence},
  author={Shafer, Glenn},
  year={2020},
  publisher={Princeton university press}
}

@article{afroogh2024trust,
  title={Trust in {AI}: progress, challenges, and future directions},
  author={Afroogh, Saleh and Akbari, Ali and Malone, Emmie and Kargar, Mohammadali and Alambeigi, Hananeh},
  journal={Humanities and Social Sciences Communications},
  volume={11},
  number={1},
  pages={1568},
  year={2024},
  publisher={Palgrave}
}

@article{donoghue2022multi,
  title={A multi-fidelity ensemble {Kalman} filter with hyperreduced reduced-order models},
  author={Donoghue, Geoff and Yano, Masayuki},
  journal={Computer Methods in Applied Mechanics and Engineering},
  volume={398},
  pages={115282},
  year={2022},
  publisher={Elsevier}
}

@article{silva2025adaptive,
  title={An adaptive hierarchical ensemble {Kalman} filter with reduced basis models},
  author={Silva, Francesco AB and Pagliantini, Cecilia and Veroy, Karen},
  journal={SIAM/ASA Journal on Uncertainty Quantification},
  volume={13},
  number={1},
  pages={140--170},
  year={2025},
  publisher={SIAM}
}

@inproceedings{hanebeck2025ensemble,
  title={Ensemble Gaussian Mixture Filter based on Projected Cram{\'e}r-von Mises Distance},
  author={Hanebeck, Uwe D and Prossel, Dominik and Popov, Andrey A and Giraldo-Grueso, Felipe and Zanetti, Renato},
  booktitle={2025 IEEE International Conference on Multisensor Fusion and Integration for Intelligent Systems (MFI)},
  pages={1--8},
  year={2025},
  organization={IEEE}
}

@article{Kikuchi_2015_ROM-EnKF,
	Author = {Ryota Kikuchi and Takashi Misaka and Shigeru Obayashi},
	Doi = {10.1088/0169-5983/47/5/051403},
	Journal = {Fluid Dynamics Research},
	Month = {sep},
	Number = {5},
	Pages = {051403},
	Publisher = {{IOP} Publishing},
	Title = {Assessment of probability density function based on {POD} reduced-order model for ensemble-based data assimilation},
	Url = {https://doi.org/10.1088%2F0169-5983%2F47%2F5%2F051403},
	Volume = {47},
	Year = 2015}

@article{Reich_2016_MLETPF,
	Author = {Gregory, A. and Cotter, C. and Reich, S.},
	Doi = {10.1137/15M1038232},
	Eprint = {https://doi.org/10.1137/15M1038232},
	Journal = {SIAM Journal on Scientific Computing},
	Number = {3},
	Pages = {A1317-A1338},
	Title = {Multilevel Ensemble Transform Particle Filtering},
	Url = {https://doi.org/10.1137/15M1038232},
	Volume = {38},
	Year = {2016}}

@article{Gregory_2017_MLETPF,
	Author = {Gregory, A. and Cotter, C.},
	Doi = {10.1137/16M1102021},
	Eprint = {https://doi.org/10.1137/16M1102021},
	Journal = {SIAM Journal on Scientific Computing},
	Number = {6},
	Pages = {A2684-A2701},
	Title = {A Seamless Multilevel Ensemble Transform Particle Filter},
	Url = {https://doi.org/10.1137/16M1102021},
	Volume = {39},
	Year = {2017}}

@techreport{Chernov_2017_MLEnKF,
	Author = {Alexey Chernov and Haakon Hoel and Kody Law and Fabio Nobile and Raul Tempone},
	Institution = {EPFL},
	Number = {22.2017},
	Title = {Multilevel ensemble {Kalman} filtering for spatially extended models},
	Type = {MATHICSE Technical Report},
	Url = {https://www.epfl.ch/labs/mathicse/wp-content/uploads/2018/10/Report-22.2017_AC_HAH_KL_FN_RT.pdf},
	Year = {2017}}

@article{Hoel_2016_MLEnKF,
	Author = {Hoel, Hakon and Law, Kody J. H. and Tempone, Raul},
	Doi = {10.1137/15M100955X},
	Journal = {SIAM Journal on Numerical Analysis},
	Number = {3},
	Title = {Multilevel ensemble {Kalman} filtering},
	Volume = {54},
	Year = {2016}}

@article{Giles_2008_MLMC,
	Author = {Michael B Giles},
	Journal = {Operations Research},
	Number = {3},
	Pages = {607--617},
	Title = {Multilevel {Monte Carlo} path simulation},
	Volume = {56},
	Year = {2008}}

@article{Giles_2015_MLMC,
	Author = {Michael B Giles},
	Journal = {Acta Numerica},
	Number = {259--328},
	Title = {Multilevel {Monte Carlo} path simulation},
	Volume = {24},
	Year = {2015}}

@article{Sandu_2016_reduced-sampling4DVar,
	Author = {A. Attia and R. Stefanescu and A. Sandu},
	Doi = {10.1002/fld.4255},
	Journal = {International Journal of Numerical Methods in Fluids},
	Number = {1},
	Pages = {28--51},
	Title = {The reduced-order hybrid {Monte-Carlo} sampling smoother},
	Url = {http://dx.doi.org/10.1002/fld.4255},
	Volume = {83},
	Year = {2016}}

@article{karpatne2024knowledge,
  title={Knowledge-guided machine learning: Current trends and future prospects},
  author={Karpatne, Anuj and Jia, Xiaowei and Kumar, Vipin},
  journal={arXiv preprint arXiv:2403.15989},
  year={2024}
}

@article{karpatne2017theory,
  title={Theory-guided data science: A new paradigm for scientific discovery from data},
  author={Karpatne, Anuj and Atluri, Gowtham and Faghmous, James H and Steinbach, Michael and Banerjee, Arindam and Ganguly, Auroop and Shekhar, Shashi and Samatova, Nagiza and Kumar, Vipin},
  journal={IEEE Transactions on knowledge and data engineering},
  volume={29},
  number={10},
  pages={2318--2331},
  year={2017},
  publisher={IEEE}
}

@article{cheng2023machine,
  title={Machine learning with data assimilation and uncertainty quantification for dynamical systems: a review},
  author={Cheng, Sibo and Quilodr{\'a}n-Casas, C{\'e}sar and Ouala, Said and Farchi, Alban and Liu, Che and Tandeo, Pierre and Fablet, Ronan and Lucor, Didier and Iooss, Bertrand and Brajard, Julien and others},
  journal={IEEE/CAA Journal of Automatica Sinica},
  volume={10},
  number={6},
  pages={1361--1387},
  year={2023},
  publisher={IEEE}
}

\end{document}